\documentclass{article}
\pdfoutput=1

\usepackage{arxiv}

\usepackage{graphicx}
\usepackage[font=footnotesize]{caption}
\usepackage[font=scriptsize]{subcaption}
\usepackage{mathtools}
\usepackage[round]{natbib}
\newcommand{\vrho}{{\boldsymbol\rho}}
\newcommand{\vc}[1]{{\bf{}#1}}
\newcommand{\mt}[1]{{\bf{}#1}}
\usepackage{xcolor}

\usepackage[utf8]{inputenc} % allow utf-8 input
\usepackage[T1]{fontenc}    % use 8-bit T1 fonts
\usepackage[implicit=false]{hyperref}% hyperlinks
\usepackage{url}            % simple URL typesetting
\usepackage{booktabs}       % professional-quality tables
\usepackage{amsfonts}       % blackboard math symbols
\usepackage{nicefrac}       % compact symbols for 1/2, etc.
\usepackage{microtype}      % microtypography

\title{Automated and accurate geometry extraction and shape optimization of 3D topology optimization results}

\author{
  M.K. Swierstra  \\
  Femto Engineering\\
  Oude Delft 137, 2611 BE Delft \\
  The Netherlands \\
  \texttt{ms@femto.eu} \\
  \And
  D.K. Gupta \thanks{During this research D.K. Gupta was affiliated with the Delft University of Technology.} \\
  University of Amsterdam\\
  Science Park 904, 1098 XH Amsterdam \\ 
  The Netherlands \\
  \texttt{d.k.gupta@uva.nl} \\
  \And
  M. Langelaar \\
  Delft University of Technology\\
  Mekelweg 2, 2628 CD Delft \\ 
  The Netherlands \\
  \texttt{m.langelaar@tudelft.nl} \\
}

\begin{document}
\maketitle

\begin{abstract}
Designs generated by density-based topology optimization (TO) exhibit jagged and/or smeared boundaries, which forms an obstacle to their integration with existing CAD tools. Addressing this problem by smoothing or manual design adjustments is time-consuming and affects the optimality of TO designs. This paper proposes a fully automated procedure to obtain unambiguous, accurate and optimized geometries from arbitrary 3D TO results. It consists of a geometry extraction stage using a level-set-based design description involving radial basis functions, followed by a shape optimization stage involving local analysis refinements near the structural boundary using the Finite Cell Method. Well-defined bounds on basis function weights ensure that sufficient sensitivity information is available throughout the shape optimization process. Our approach results in highly smooth and accurate optimized geometries, and its effectiveness is illustrated by 2D and 3D examples.
\end{abstract}

% keywords can be removed

\section{Introduction} \label{intro}
\begin{figure*}[t]
	\centering
	\begin{minipage}{\textwidth}
		\centering
		\includegraphics[width=\textwidth]{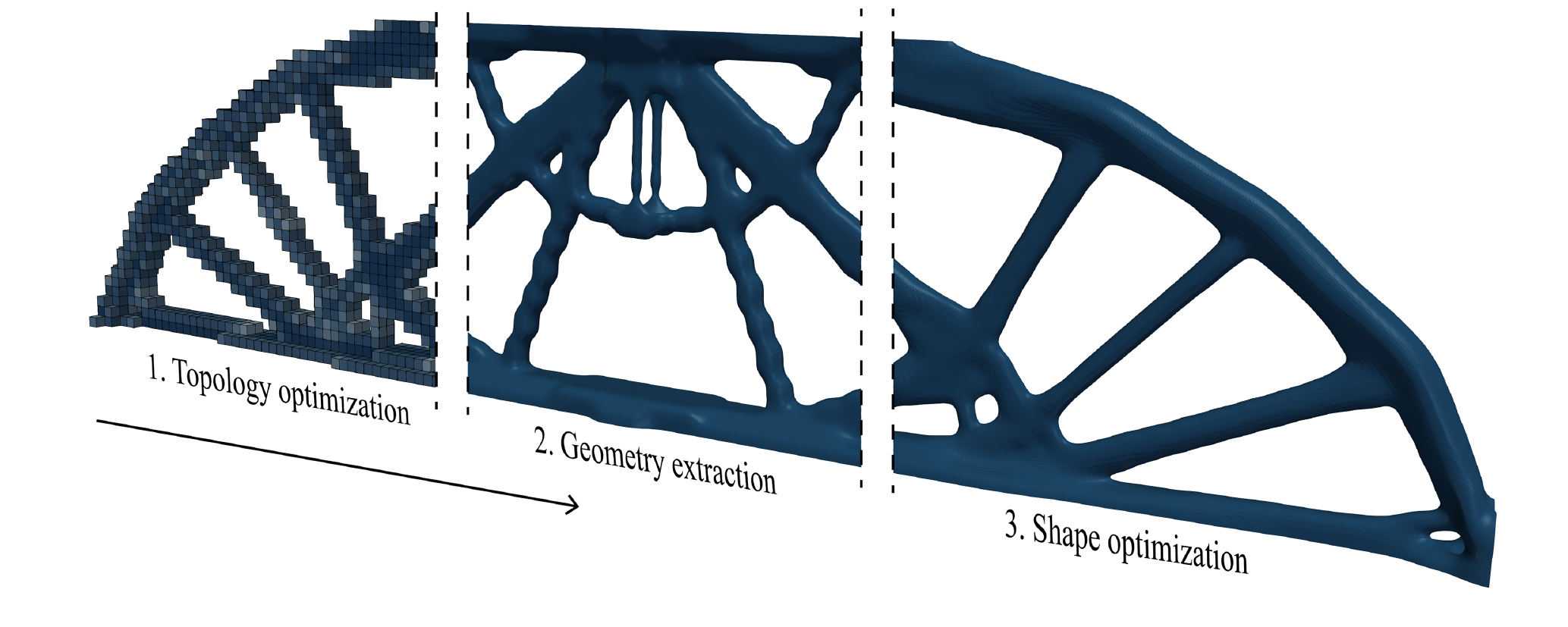}
		\captionsetup{width=\textwidth}
		\caption{Three stage structural design optimization process: topology optimization, geometry extraction and shape optimization. This paper focusses on the post-processing of TO results, i.e. Stages 2 and 3.}
		\label{fig:3-staged_process}
	\end{minipage}
\end{figure*}

% Motivation
Topology optimization (TO) is an increasingly popular generative design method that forms an established part of the design process in various branches of industry. The density-based approach (\citet{bendsoe2003sigmund}) is dominant and available in several commercial software packages. TO is most often used to generate design concepts in an early stage of the design process, and optimizes a material distribution defined in terms of local density variables. Typical density-based TO results, in combination with common filtering techniques, exhibit intermediate densities representing virtual semi-dense material, and jagged boundaries appear due to the use of a finite-element-based design discretization (e.g. Figures \ref{fig:3-staged_process}(left) and \ref{fig:MBB_beam_stage1}). In addition, the analysis accuracy is typically low in the TO process, due to the inaccurate representation of a smooth structural boundary and the low-order finite elements typically used. Currently, TO results are subsequently redrawn or post-processed manually for further design iterations and higher-fidelity analysis. Given the present efficient TO processes, this manual post-processing step increasingly becomes a bottleneck time-wise and also the optimality of the TO design is lost. In addition, a more seamless connection between TO and existing CAD tools for e.g. design validation is desired.

%Previously other (high-resolution)
How to bridge the gap between TO and CAD is a longstanding challenge (e.g. \citet{olhoff1991cad}). Various methods have been proposed to obtain smooth, high-resolution TO results in order to reduce the aforementioned gap. \citet{costa2003layout}, \citet{stainko2006adaptive} and \citet{nana2016towards} used $h$-refinement to improve the boundary resolution and to reduce the number of design variables by clearing large void areas. \citet{nguyen2010computational} decoupled the structural analysis mesh and the density design variables to obtain a higher resolution TO result at limited computational cost. \citet{kumar1996synthesis} used a shape density function, interpolated from nodal density variables, to optimize both topology and shape at once. Similarly, \citet{kang2012nodal} mapped the nodal density variables onto the material design space using Shepard interpolants, enabling a more complex spatial distribution of the material density within an individual element. Additionally, \citet{wang2013adaptive} adaptively refined the density field by adding nodal density variables without remeshing the finite element model, while \citet{Wang2014adaptive} extended this concept with independent adaptive refinement of the analysis mesh.

The aforementioned refinement methods inevitably increase the computational cost of the TO process considerably, in particular in 3D. Rather than improving the TO result, \citet{papalambros1990integrated} proposed a three-stage process for structural design optimization: 1) generate a TO result, 2) process the density distribution to obtain a practical and realistic structure, and 3) refine the final topology by detailed shape optimization. Our proposed approach essentially also follows these three steps, as illustrated in Figure \ref{fig:3-staged_process}. A shape optimization is needed because the design obtained using TO loses its optimality or even feasibility as soon as it is translated into a crisp and smooth geometry. This is caused by thresholding of intermediate densities to obtain a binary image and arbitrary smoothing of the boundary. A subsequent shape optimization of the obtained geometry is used to retrieve a feasible and optimized design.

Various implementations of this general philosophy have been proposed previously. \citet{maute1995adaptive} proposed two methods, one involving the previously mentioned $h$-refinement, and the other a three-stage approach. The TO result was mapped to a spline-based geometry description, which was subsequently remeshed for another TO cycle. These cycles were repeated until the TO result became smooth and crisp. \citet{bremicker1991integrated} processed the TO result into a binary image from which they obtained a truss or continuum structure. \citet{lin2000automated} also firstly obtained a binary image after which they described the external boundaries using B-splines and the internal features by standard shapes. More recently, \citet{yi2017identifying} developed a similar method in which geometric features of the TO result were detected to construct a parametrized CAD model. \citet{tang2001integration} and \citet{chang2001integration} extracted a geometry by averaging the obtained boundary nodes of a thresholded TO result and applied a least-square fitting using B-splines. \citet{hsu2001interpreting} fitted B-splines on density contours to interpret the TO result.

Once obtained, a spline representation of a geometry can be used for a subsequent shape optimization. Spline representations can be obtained in 2D but automatically creating spline surfaces to represent a complex 3D TO geometry is challenging. \citet{hsu2005interpreting} attempted to overcome this difficulty by sweeping 2D contours into a 3D surface. Also commercial software is available to perform this step in an interactive manner, however user input remains necessary. Furthermore, spline representations add complexity, as the transition from TO model to shape optimization model involves remeshing, and the change in design parametrization typically requires a different sensitivity analysis implementation. Most importantly, in previous works no method thus far has been fully satisfactorily demonstrated in 3D, yet the vast majority of TO cases considered in industry are three dimensional.

% Goal
This paper focuses on the automated post-processing of both 2D and 3D density-based TO results in order to form a convenient bridge to other CAD tools. The aim is to obtain a structural design optimization process capable of generating optimized, smooth and crisp geometries with accurate, optimized performance, without any manual labour. This includes creating a mesh only once for the TO, and utilizing sensitivity analysis procedures that remain essentially the same throughout the structural design optimization process. To this end, we propose a fully integrated level-set-based shape optimization, where the initial level-set is constructed from the result of a density-based TO process. The end result will be an optimized geometry defined by a level-set function, which can be post-processed into other geometry representations. Note that generation of e.g. CAD-compatible NURBS descriptions is a very different challenge that falls outside the scope of this paper.

The novelty of our approach lies in Stage 2 and 3 of the design process shown in Figure \ref{fig:3-staged_process}: seamlessly combining geometry extraction and shape optimization of density-based TO results while satisfying the design constraints after each stage. To achieve this, we employ the novel combination of a parametric level-set formulation with bounded level-set gradients and the Finite Cell Method (FCM) to increase analysis accuracy. While building on these existing techniques, part of the novelty of this work lies in their specific combination, which has not been reported before in the literature. The level-set gradients are rigorously controlled to prevent convergence problems, by defining a new relation between the maximum level-set gradient and the weights of the parametrization functions. Our approach has been conceived with the aim to be easy to apply both in 2D and 3D, while most reported methods have only been demonstrated for 2D examples.

An alternative approach, to use a level-set method (LSM) from the start of the design optimization process, is not our preferred choice for three reasons. All commercial software implementations of TO use the density-based setting, so to maximize the applicability of our proposed method we choose this as our starting point. Furthermore, in general, LSMs show stronger dependence on the chosen initial design compared to density-based TO (\citet{vanDijk2013level}). The use of topological derivatives improves this aspect but these are not straightforward to formulate and differ strongly for different kinds of optimization objectives and constraints. Thirdly, we employ FCM-based local refinement to increase the analysis accuracy in the level-set-based final shape optimization. This is affordable when boundary motion remains modest, but for the full TO process it would come at high computational cost due to frequent refinement updates. Thus, instead of opting for a single method, in our view combining the strengths of density- and level-set-based approaches is most advantageous.

%Outline
The following three sections discuss our approach to each of the three stages of the structural design optimization process. To illustrate the process, 2D-cases are optimized, with a focus on the common problem of minimizing compliance subject to a volume fraction constraint. Section 5 discusses the performance of the proposed method using both 2D and 3D compliance minimization case studies. Finally, in Section 6, conclusions and recommendations are given.

\section{Stage 1 - Topology optimization}
The first stage of the design optimization process is the topology optimization. The popular density-based SIMP method was independently introduced by \citet{bendsoe1989optimal} and \citet{zhou1991coc}. In the SIMP method one defines a density in each element of the domain as a design variable that can vary between 0 (void) and 1 (solid). The stiffness properties of an element are linked to this density using the following power law:
\begin{equation}\label{eq:SIMP}
E = \rho^p \cdot E_0
\end{equation}
where $p$ is the penalization parameter and $E_0$ is Young's modulus for the isotropic material. Equation \ref{eq:SIMP} penalizes intermediate densities as these provide relatively low stiffness.
In this paper we base the numerical examples on the well-known compliance minimization problem to illustrate the proposed post-processing method. This optimization problem is defined as:
\begin{equation}
\begin{array}{lll}
&\underset{\vrho}{\min} &C = \vc{f}^T\vc{u} \label{eq:compliance}\\
&\textrm{s.t.}    &\mt{K}(\vrho)\vc{u}=\vc{f},\\
&                 &V(\vrho)/V_{\max}-1\leq 0, \\
&                 &0<\rho_{\min} \leq \rho_i \leq 1 \qquad i=1\ldots{}n. \\
\end{array}
\end{equation}
Here $C$ denotes the compliance objective, and $\vrho$, $\vc{u}$ and $\vc{f}$ are respectively the density design variable vector, displacement and load vector. Furthermore $\mt{K}$ represents the FE stiffness matrix, where the SIMP interpolation is used to define the dependence on $\vrho$. $V$ and $V_{\max}$ denote the actual and allowed volume, and the last line defines the bounds on the $n$ design variables that define the material layout. For further discussion of this classical problem, we refer to e.g. \citet{bendsoe2003sigmund}.

\begin{figure}[h]
	\centering
	\includegraphics[width=0.6\textwidth]{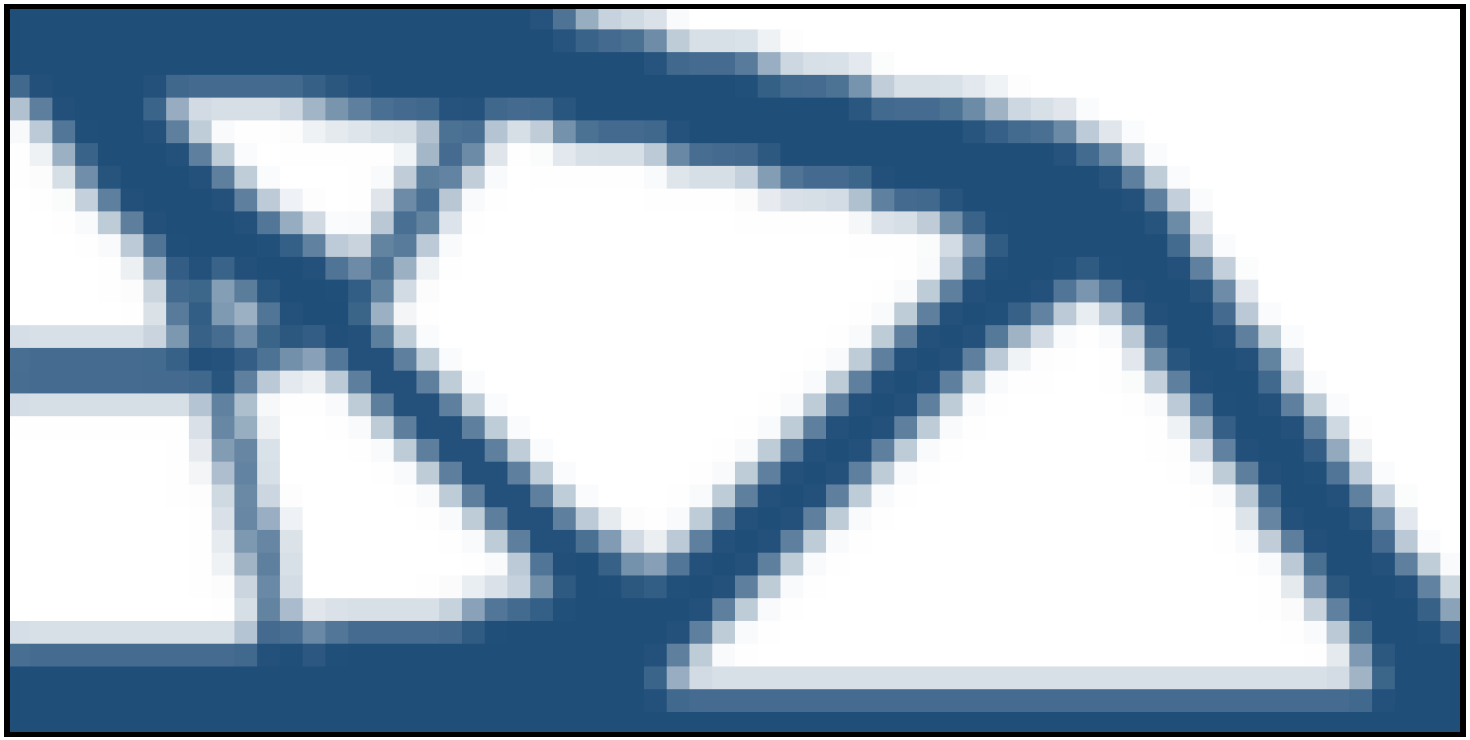}
	\captionsetup{width=0.6\textwidth}
	\caption{Example of an topology-optimized MBB beam on a 64$\times$32 grid using bilinear shape functions. A coarse grid is chosen to clearly illustrate typical artifacts.}
	\label{fig:MBB_beam_stage1}
\end{figure}

A typical TO result for this problem is shown in Figure \ref{fig:MBB_beam_stage1}, obtained using the Python version of the 99-line Matlab code of \citet{sigmund200199}. Two features appear which make TO results not directly usable in a CAD setting. First of all, jagged boundaries appear due to the use of element-wise constant densities as a geometry description. Secondly, the result is not fully discrete because some elements have an intermediate density. In this example a coarse mesh has been chosen deliberately to highlight these aspects --- in practice, the designer is free to choose the mesh appropriate for the desired level of detail and computational cost. Mesh refinement reduces the severity of the aforementioned problems, but also leads to a rapid increase in computational cost, particularly in 3D. Moreover, at the finest length scale of the generated design these problems can invariably be observed.

\section{Stage 2 - Geometry extraction}
Density-based TO results exhibit jagged and blurred boundaries, while designers and established CAD tools require smooth and clear-cut geometry descriptions. Smooth boundaries can be represented using different methods. In this work, a level-set function (LSF) is used as the geometry description because of its inherently smooth characteristics. Furthermore, an LSF is relatively easy to extend to 3D compared to for example spline representations and a parametric LSF makes it possible to utilize the same sensitivity analysis as used for the TO stage.

\subsection{Density field to LSF}\label{sec:rho2LSF}
The design represented by the optimized density distribution from the SIMP method needs to be translated into a geometry description based on an LSF. Our LSF is described by a summation of Radial Basis Functions (RBFs), similar to \citet{luo2008level}, each located at the centroid $\mathbf{x}_i$ of an element $i$. Gaussian RBFs $N_i$ will be used and are described by:
\begin{equation}\label{eq:Ni}
N_i(\mathbf{x}) = e^{-\left(\frac{R_i(\mathbf{x})}{h}\right)^2}, \qquad\mathrm{with}\quad  R_i(\mathbf{x})=||\mathbf{x}_i-\mathbf{x}||_2 .
\end{equation}
Here $R_i$ is the radial distance from the location of the RBF and $h$ is the element edge length. The element edge length is chosen as 1 in this research, so $h$ is omitted in further expressions related to the RBFs. Each RBF is multiplied with a certain weight $w_i$, which form the design variables in a subsequent shape optimization. The LSF $\phi(\textbf{x},\textbf{w})$ is the summation of the RBFs in the design domain and thus becomes:
\begin{equation}\label{eq:phi}
\phi(\textbf{x},\mathbf{w}) = \left(\sum_{i=1}^{n}{e^{-R_i^2}\cdot w_i}\right)  - \theta,
\end{equation}
where $n$ is the number of elements, and a constant $\theta$ is subtracted to shift the LSF globally. Initially, we choose $\theta=0$. Weights $w_i$ control the shape of the LSF, which defines the boundary of the structure. These weights will also be used as design variables in Stage 3. In the following, the dependencies of $\phi$ will be omitted for brevity.

To initialize the weights, a set of linear equations can be set up to determine the weights of the RBFs, such that the LSF matches the densities obtained at Stage 1 in all element centroids:
\begin{equation}\label{eq:stage2_LSF}
\mathbf{\Phi} \mathbf{w} = \boldsymbol{\rho},
\end{equation}
where
\begin{equation}\label{eq:Phi}
\Phi_{ij} = e^{-R_{ij}^2}, \qquad\mathrm{with}\quad
R_{ij} = || \mathbf{x}_i - \mathbf{x}_j ||_2 ,
\end{equation}
and where $\mathbf{x}_i$ and $\mathbf{x}_j$ denote respectively the centroids of elements $i$ and $j$. While Gaussian RBFs theoretically have infinite support regions, for numerical convenience a finite support radius is considered as the RBF quickly decays to zero. The RBFs used in this research have a support diameter of 7 elements. As a result, matrix $\mathbf{\Phi}$ is sparse and Equation~\ref{eq:stage2_LSF} can be solved with limited effort.

\subsection{Thresholding the LSF}
It is common in level-set-based TO methods that a function $H(\phi)$ is used to describe the structural domain and the void domain, where $H(\phi>0)$ is the interior of the structure, and $H(\phi<0)$ is the outside region. To allow for gradient-based optimization, a continuous Heaviside approximation is used:
\begin{equation}\label{eq:Heaviside}
H(\phi) = \frac{1}{1+e^{-\kappa \phi}}.
\end{equation}
For further analysis, it is convenient to again define a density field, linked to the LSF. This density field is denoted by $\hat{\rho}$, and is given by:
\begin{equation}\label{eq:newrho}
\hat{\rho}=\rho_0 + (1-\rho_0)H(\phi).
\end{equation}
Equation~\ref{eq:newrho} links negative LSF-values to void regions (with density $\rho_0$, a minimum density value used for numerical stability) and positive LSF-values to solid regions (with density 1.0). In this way, the intermediate densities observed in the design obtained from Stage 1 can be minimized in the extracted geometry. A finite value of $\kappa$ maintains continuity of the geometry description, which allows for subsequent gradient-based shape optimization. A discussion on sensitivity analysis follows in Section \ref{sec:sens_analysis}.

\begin{figure}[h]
	\centering
	\includegraphics[width=0.8\textwidth]{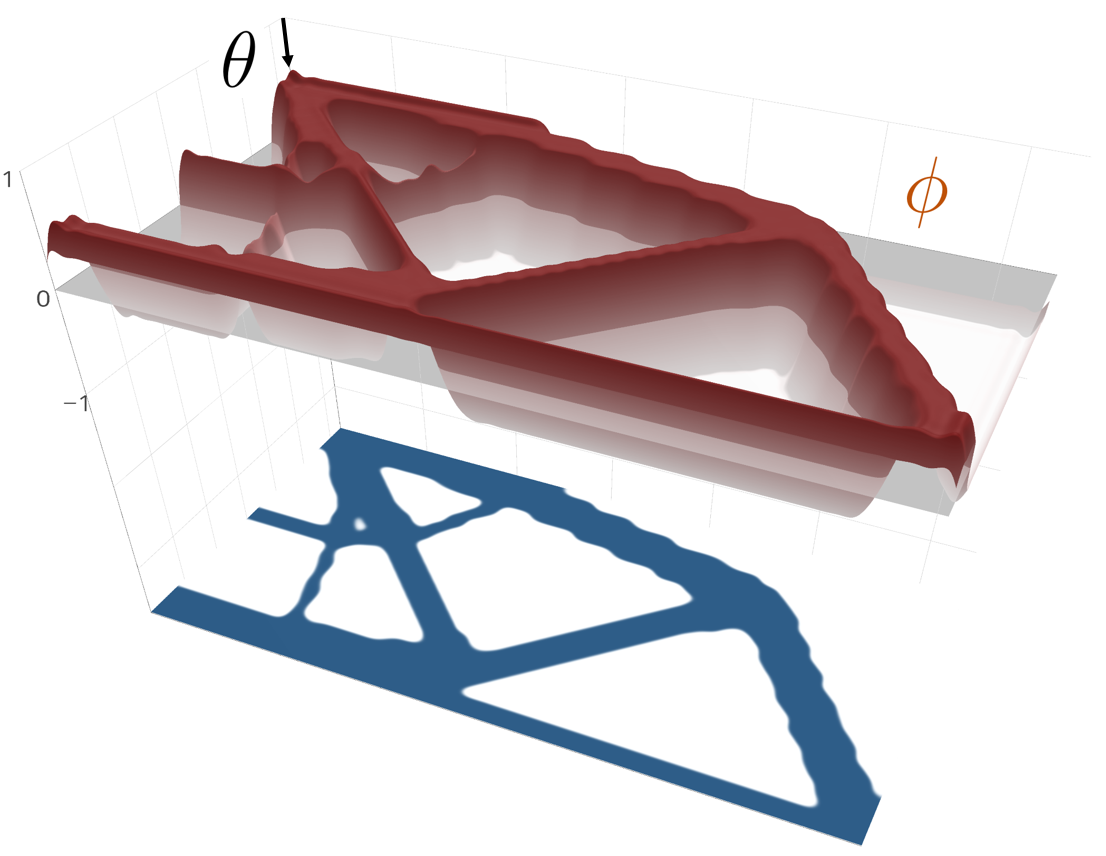}
	\captionsetup{width=0.8\textwidth}
	\caption{Projection of the shifted LSF $\phi$ (red) to a geometry (blue) using the smooth Heaviside function of Equation \ref{eq:Heaviside}. The image shows the LSF and geometry extracted from the SIMP example shown in Figure \ref{fig:MBB_beam_stage1}.}
	\label{fig:LSF_projected}
\end{figure}

The LSF obtained by solving Equation \ref{eq:stage2_LSF}, $\phi_0$, is fully positive because the element densities of the TO result are as well. Using Equation \ref{eq:Heaviside} would then result in a completely solid design domain. So, the LSF should be lowered, to an extent that the area within the zero-level contour satisfies the desired volume fraction, see Figure \ref{fig:LSF_projected}. This is equivalent to setting the appropriate threshold value. The shift value $\theta$ of the LSF must satisfy Equation \ref{eq:stage2_thres}:
\begin{equation}\label{eq:stage2_thres}
\frac{\int_{\Omega}{ H(\phi_0-\theta) d\Omega}}{\int_\Omega d\Omega}-V_{f}=0 ,
\end{equation}
where $\phi=\phi_0-\theta$ and $V_{f}$ is the set volume fraction of the optimization problem. Numerical integration is performed using the Gauss integration points from Stage 1. Shift value $\theta$ is found by solving the nonlinear single-variable problem given by Equation \ref{eq:stage2_thres} using the Newton-Raphson method, and is kept constant in the remainder of the procedure.

\section{Stage 3 - Shape optimization}
The weights of the RBFs $w_i$, obtained in Stage 2, can be used as design variables for a subsequent shape optimization of the smooth geometry, similar to e.g. \citet{luo2009level}. The aim of this optimization process is to finetune the shape of the design, however for added flexibility the chosen formulation also allows for topological changes. This has the advantage that e.g. inefficient members or structural details can be removed in this stage. Nevertheless, the main steps in defining the structure's topology are taken in Stage 1, while in Stage 3 the focus is on refining its shape. In principle the optimization problems considered in stages 1 and 3 can differ (e.g. inclusion of additional constraints), however in this paper we keep them the same. Hence here a problem similar to Equation~\ref{eq:compliance} is optimized, except that instead of density variables $\vrho$ the weights $\vc{w}$ form the design vector. RBF weights $\bf{w}$ define the LSF $\phi$, which in turn defines the density $\rho$ in every integration point in the domain.

A structural and sensitivity analysis are needed in order to perform a shape optimization using an LSM. The slope of the LSF should be controlled as well to ensure good optimization convergence (\citet{vanDijk2013level}). The following subsections discuss how these aspects are treated in the proposed method.

\subsection{Structural analysis}
The structural analysis of the geometry defined implicitly by the LSF is not straightforward. The original TO mesh and the LSF boundary do not match. Remeshing is an option but this is considered less robust, especially without manual intervention, and can negatively affect the consistency of design sensitivity information. We choose to instead perform the structural analysis using \textit{p}-FEM (\citet{Szabo1991Finite}) in combination with an adaptive numerical integration method. \textit{p}-FEM allows for the use of the same grid as used for the TO and its accuracy is improved by increasing the polynomial order of the FE-interpolation functions. This is an established modeling approach, and further details can be found in e.g. \citet{Babuska1981}.

\begin{figure}[h]
	\centering
	\includegraphics[width=0.7\textwidth]{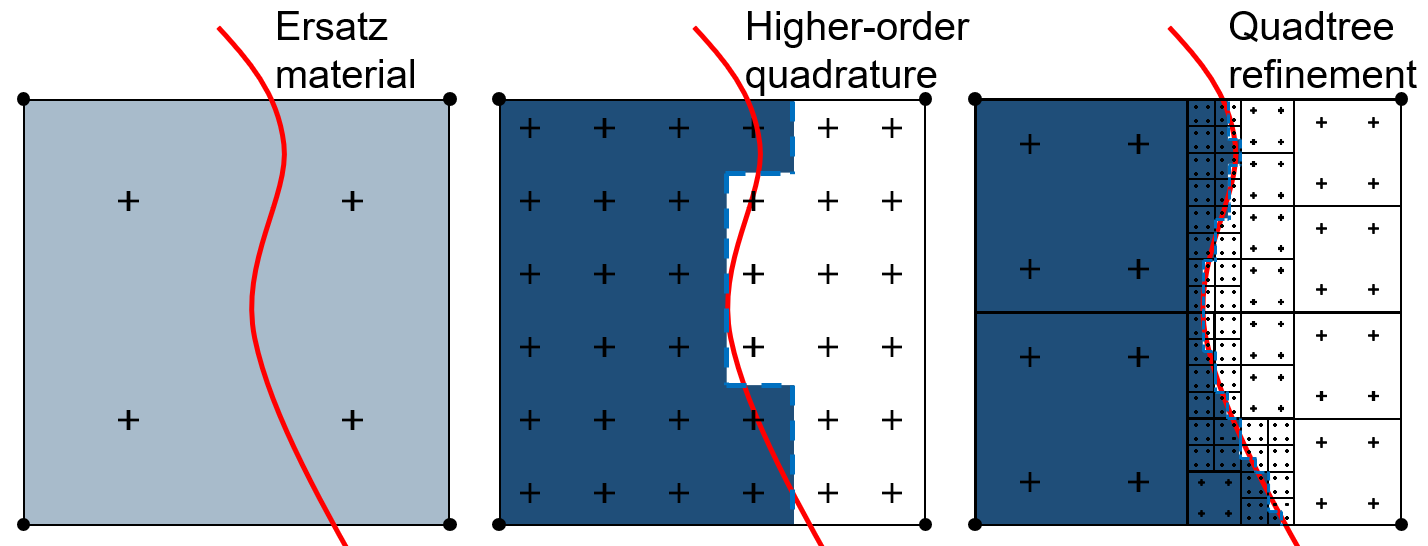}
	\captionsetup{width=0.7\textwidth}
	\caption{Examples of adaptive numerical integration methods: `Ersatz material' by e.g. \citet{allaire2004structural}, higher-order quadrature by e.g. \citet{parvizian2007finite} and a quadtree refinement by e.g. \citet{Abedian2013performance}. The \mbox{`+'-signs} indicate integration points.}
	\label{fig:adapt_int}
\end{figure}

Several methods exist to perform the adaptive numerical integration, see Figure \ref{fig:adapt_int}. \citet{Abedian2013performance} showed that a quadtree refinement is a relatively efficient approach to gain accurate results, so this method is used for the adaptive numerical integration. In 3D, an equivalent octree refinement scheme can be applied. This combination of methods, \textit{p}-FEM and an adaptive integration method, is introduced by \citet{parvizian2007finite} under the name Finite Cell Method (FCM). FCM captures both the geometry and the response of the structure more accurately than the initial low-order FE-model. In this way, the same grid used in Stage 1 can be used while the geometry is described using an LSF. The obtained accuracy is verified in Section \ref{sec:performance} for the performed case studies. For details on the FCM formulation, we refer to the original work by \citet{parvizian2007finite}.

\begin{figure}[h]
	\centering
	\includegraphics[width=0.6\textwidth]{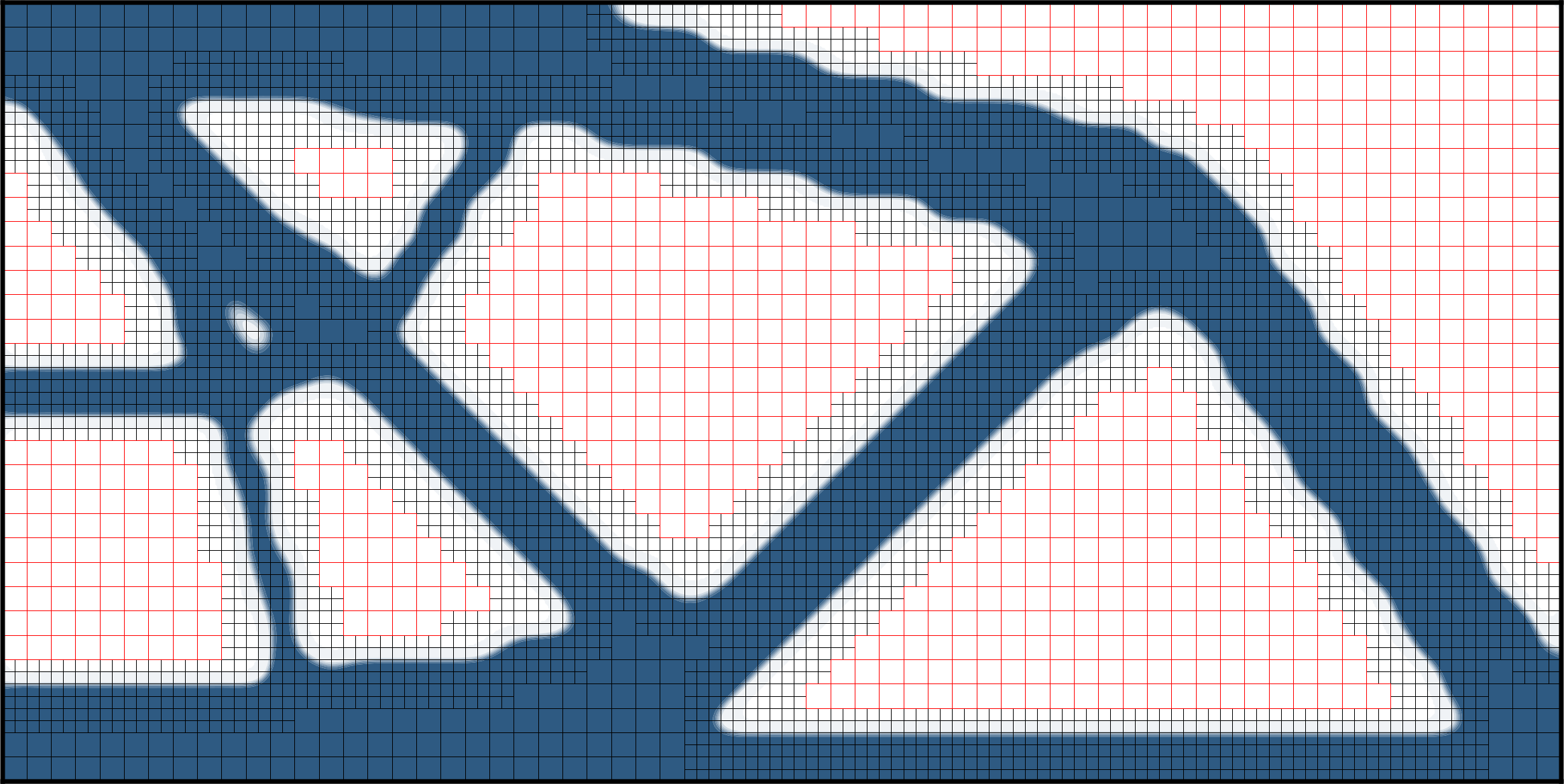}
	\captionsetup{width=0.6\textwidth}
	\caption{FCM grid on the extracted geometry of the MBB beam of Figure \ref{fig:LSF_projected} using one level of quadtree refinement. Red squares correspond to discarded elements and black squares to integration cells.}
	\label{fig:FCM_grid_MBB}
\end{figure}

The quadtree refinement is extended further from the boundary by approximately one element, see Figure \ref{fig:FCM_grid_MBB}. This allows the boundary to move during the shape optimization while remaining within the accurate quadtree integration band, while avoiding frequent updating of the integration datastructure during optimization. As the geometry is the result of an earlier TO process (Stage 1), boundary motion in this final shape optimization stage is typically small. Note also that no remeshing is required, only the integration cells are redefined. Elements at further distances from the structural domain are discarded.

To analyze the design defined by the LSF using FCM, the densities $\hat{\rho}$ (Equation \ref{eq:newrho}) are evaluated in the integration points. The element stiffness matrix and subsequently the global stiffness matrix can then be assembled using the SIMP material interpolation in each integration point.

\subsection{Sensitivity analysis}\label{sec:sens_analysis}
The sensitivity analysis for Stage 3 is very similar to Stage 1. The derivative of the objective (e.g. compliance $C$) with respect to the design variables (RBF weights $\mathbf{w}$) is desired. The chain rule of differentiation gives:
\begin{equation}\label{eq:dC_ds}
\frac{\partial C}{\partial w_i} = \frac{\partial C}{\partial \hat{\rho}_k} \frac{\partial \hat{\rho}_k}{\partial \phi_k} \frac{\partial \phi_k}{\partial w_i} ,
\end{equation}
where a summation convention applies to index $k$, and $\rho_k$ and $\phi_k$ denote the density and LSF-value at a particular integration point $\mathbf{x}_k$, respectively.

From Equations \ref{eq:Ni} and \ref{eq:phi}, it follows that the derivative of the LSF $\partial\phi_k/\partial{}w_i$ equals $N_i$. Note that this derivative is similar to the matrix $\mathbf{\Phi}$ in Equation \ref{eq:Phi}, except the centroid location $\mathbf{x}_j$ of an element $j$ is replaced by the location $\mathbf{x}_k$ of an integration point $k$.
As already noted in Section~\ref{sec:rho2LSF}, the finite support of $N$ renders $\partial\phi/\partial\mathbf{w}$ a sparse operator.
From Equation \ref{eq:newrho}, relating the LSF in each integration point to the density, the second derivative in Equation \ref{eq:dC_ds} becomes:
\begin{equation}\label{eq:drho_dphi}
\frac{\partial \hat{\rho}_k}{\partial \phi_k} = (1-\rho_0) \frac{\kappa e^{-\kappa \phi_k}}{(1+e^{-\kappa \phi_k})^2} ,
\end{equation}
where the definition of the smooth Heaviside (Equation~\ref{eq:Heaviside}) has also been used.

Finally, the partial derivative of the compliance with respect to the density in an integration point $k$ should be determined. This is the same sensitivity as computed in the SIMP formulation applied in Stage 1 (\citet{bendsoe2003sigmund}):
\begin{equation}
\frac{\partial C}{\partial \hat{\rho}_k} = -p\hat{\rho}_k^{p-1}\textbf{u}_e^T\textbf{K}_{e,k}\textbf{u}_e ,
\end{equation}
where $\textbf{u}_e$ is the displacement vector at element level, and $\textbf{K}_{e,k}$ the contribution to the element stiffness matrix at integration point $k$. The derivative of Equation \ref{eq:dC_ds} can be split into two parts: the first term is similar to that of Stage 1, while the second and third terms are problem-independent and solely related to the LSM. Since our focus is on the post-processing procedure itself, the proposed design optimization process has only been implemented and tested here for minimum compliance problems. However, because of this independence it applies in a similar fashion to other types of optimization problems.

\subsection{LSF slope control}\label{sec:LSF_control}
In the sensitivity expression discussed above (Equation~\ref{eq:dC_ds}), the derivative of the density with respect to the LSF, $\partial\hat{\rho}/\partial\phi$, requires careful treatment. This derivative becomes zero in the entire domain for high values of $\kappa\phi$ in Equation \ref{eq:Heaviside}, i.e. for a very steep Heaviside function, except for the structural boundary where it approaches infinity. This extreme spatial variation and localization of sensitivity information results in slow shape changes and convergence problems \citet{vanDijk2013level}. To counteract this, the LSF $\phi$ and its spatial derivative $\nabla\phi$ should be controlled, to be able to use a fixed value for $\kappa$. To avoid steep gradients of the LSF, different approaches exist in the literature. For conventional LSMs, regular signed-distance reinitialization is common, e.g. \citet{allaire2004structural}. To avoid the cost and inaccuracy associated to reinitialization, others have proposed adding an artificial regularization energy term to the objective (\citet{zhu2015structural}). Setting the required strength of this term is however not trivial. In the context of parametric level-set approaches, as used in this paper, previous studies have used bounds on the RBF weights (\citet{pingen2010parametric}) but so far these have been chosen arbitrarily. As part of the proposed method, we here present a new mathematical relation between the maximum LSF slope and the bounds on the weights, to rigorously control the LSF steepness. Our approach is to determine the steepest LSF slope that can occur for a given bound on the RBF weights, and use this insight to choose steepness parameter $\kappa$ to ensure sufficient sensitivity magnitudes near the structural boundary.

The maximum value that an LSF can take, can be related to the maximum weight $w_{max}$ of the RBFs. An 1D-RBF located at $x_0$ and evaluated at $x$, is given by:
\begin{equation}
\phi = e^{-(x-x_0)^2} \cdot w_{max} ,
\end{equation}
and its derivative reads:
\begin{equation}
\frac{d\phi}{dx} = -2(x-x_0) \cdot e^{-(x-x_0)^2} \cdot w_{max} .
\end{equation}
An infinite summation of RBFs located at integer $x$-locations $x_i$, evaluated at $x=0$, is numerically found to be:
\begin{equation}
\phi_{max} = \sum_{x_i=-\infty}^{\infty}{e^{-(0-x_i)^2} \cdot w_{max}} = \vartheta_3(0,e^{-1}) \cdot w_{max} \approx 1.7726 \cdot w_{max},
\end{equation}
where $\vartheta_3$ is the elliptic theta function. The maximum LSF-value for a higher dimensional LSF is simply:
\begin{equation}\label{eq:LSFmax}
\phi_{max} = 1.7726^D \cdot w_{max},
\end{equation}
as found by numerical validation, where $D$ is the dimension of the design (e.g. $D=2$ for a 2D-case).

\begin{figure}[h]
	\centering
	\includegraphics[width=0.625\textwidth]{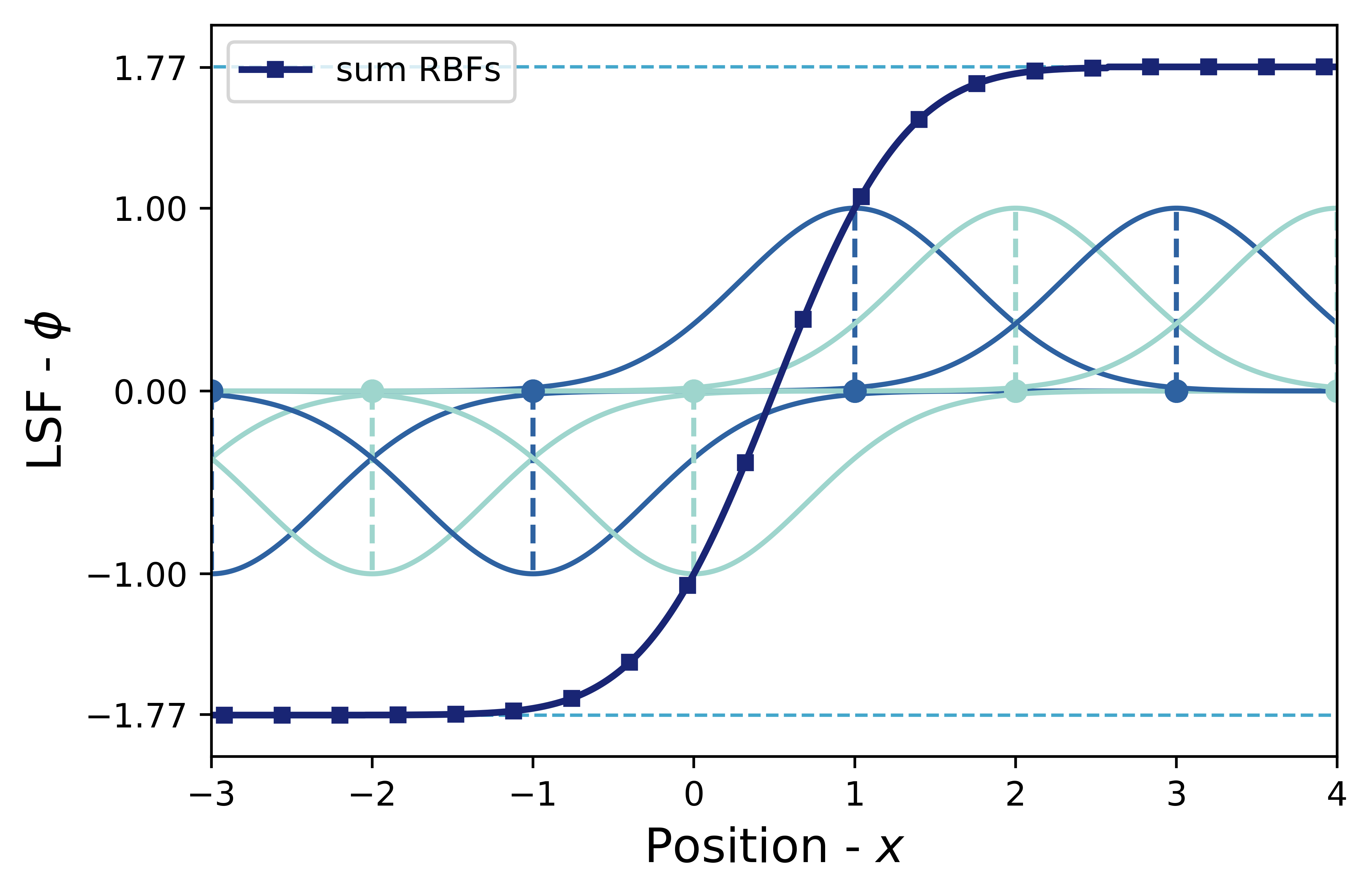}
	\captionsetup{width=0.625\textwidth}
	\caption{Infinite summation of 1D RBFs located at integer $x$-locations $x_i$. RBF weights are set to $w=\mathrm{sign}(x-0.5)$, i.e. $w=-1$ for RBFs located at $x_i<0.5$, and $w=1$ for those located at $x_i>0.5$.}
	\label{fig:RBFsum1D}
\end{figure}

Considering an extreme weight distribution with $w=\mathrm{sign}(x-0.5)w_{max}$, the maximum possible slope occurs at the transition from the minimum LSF-value to the maximum one, as depicted in Figure \ref{fig:RBFsum1D}. The summation of the derivatives of the RBFs is:
\begin{equation}\label{eq:LSFgradmax}
\left( \frac{d\phi}{dx} \right)_{max} = 4\sum_{x_i=1}^{\infty}{(x-x_i) \cdot e^{-(x-x_i)^2} \cdot w_{max}} \approx 2.2094 \cdot w_{max},
\end{equation}
where $x=0.5$, because this is where the maximum slope occurs (i.e. right in between the transition from negative to positive as can be observed in Figure \ref{fig:RBFsum1D}).

\begin{figure}[h]
	\centering
	\includegraphics[width=0.7\textwidth]{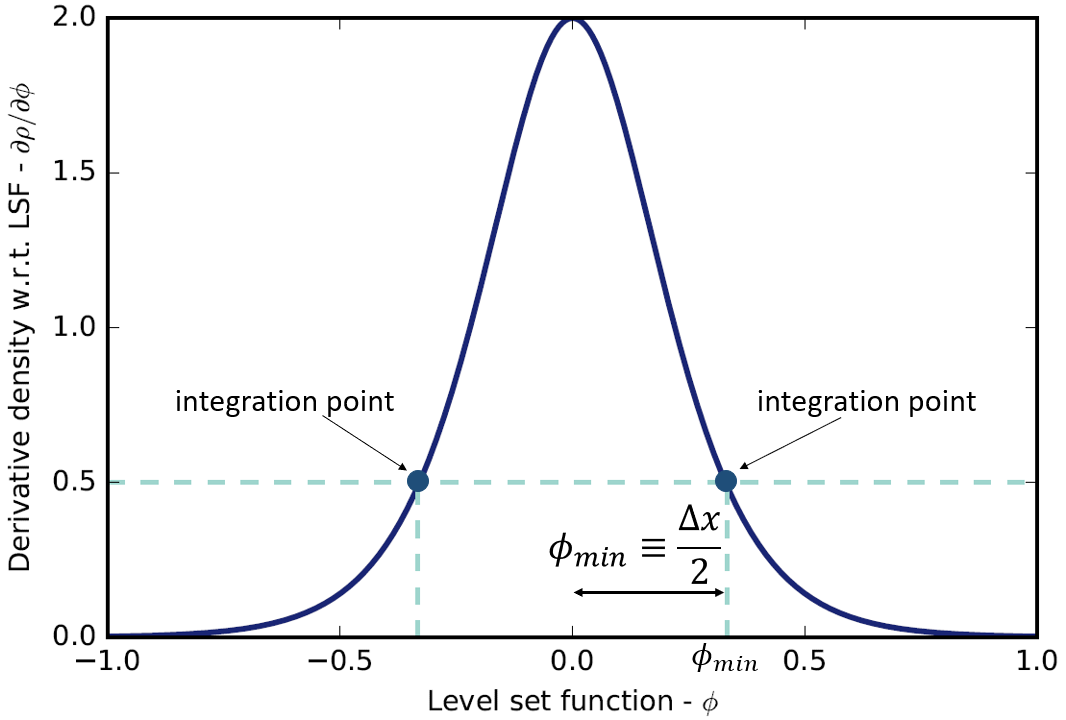}
	\captionsetup{width=0.7\textwidth}
	\caption{Illustration of determining the steepness of the Heaviside function based on the maximum distance between two integration points nearest to the boundary ($\phi=0$).}
	\label{fig:calc_kappa}
\end{figure}

The maximum slope scales linearly with the maximum value of an RBF summation. Using this linear relationship between Equations \ref{eq:LSFmax} and \ref{eq:LSFgradmax}, results in the following general expression for the maximum gradient magnitude of an LSF bounded by the weights $w_{max}$:
\begin{equation}\label{eq:max_slop_phi}
|\nabla \phi|_{max} = 1.2464 \cdot 1.7726^{D} \cdot w_{max}.
\end{equation}
Next, a suitable value for $\kappa$, which controls the steepness of the Heaviside approximation (Equation~\ref{eq:Heaviside}), can be determined using the maximum possible value $|\nabla \phi|_{max}$ of Equation \ref{eq:max_slop_phi} and the average distance between integration points,
\begin{equation}
\Delta x = \frac{2^{-qt}}{P+1} ,
\end{equation}
where $qt$ is the number of quadtree levels and $P$ the $p$-FEM shape function order, which sets the amount of Gauss points inside an integration cell. The maximum distance between the boundary and the nearest integration point is $\Delta x/2$.
The most critical situation occurs when the boundary $(\phi=0$) falls exactly between two integration points (Equation \ref{fig:calc_kappa}). From this we can estimate the following upper bound on the corresponding absolute LSF-values in these integration points:
\begin{equation}\label{eq:phi_ip}
|\phi_{ip}| = |\nabla \phi|_{max} \cdot \frac{\Delta x}{2} .
\end{equation}
In these points, the sensitivity $\partial\hat{\rho}/\partial{\phi}$ should not drop below a certain minimum magnitude. Substitution of $|\phi_{ip}|$ in Equation \ref{eq:drho_dphi} gives:
\begin{equation}\label{eq:drho_dphi2}
\left.\frac{\partial \hat{\rho}}{\partial \phi}\right|_{min} = (1-\rho_0)  \frac{\kappa \cdot e^{-\kappa |\phi_{ip}|}}{(1+e^{-\kappa |\phi_{ip}|})^2} ,
\end{equation}
which can be numerically solved for $\kappa$ when setting a chosen minimum value for the derivative in an integration point. This minimum value for the derivative should not be close to zero because then there would be no sensitivities defined around the boundary. A too large value would result in a boundary region containing an unnecessary large amount of intermediate densities, as $\kappa$ controls the width of the transition zone between solid and void region (Equation \ref{eq:Heaviside}). Between these extremes, many values are suited to limit the gradient magnitude. In this research a value of 0.5 is chosen as the minimum value for the derivative of the density with respect to the LSF. This value in combination with $P=2$ and $qt=1$, results in $\kappa \approx 25$. Figure~\ref{fig:calc_kappa} illustrates the procedure to determine a fixed value for $\kappa$.
\clearpage
\section{Case studies}
\begin{figure}[h]
	\centering
	\begin{minipage}[t]{.4\textwidth}
		\centering
		\includegraphics[width=\linewidth]{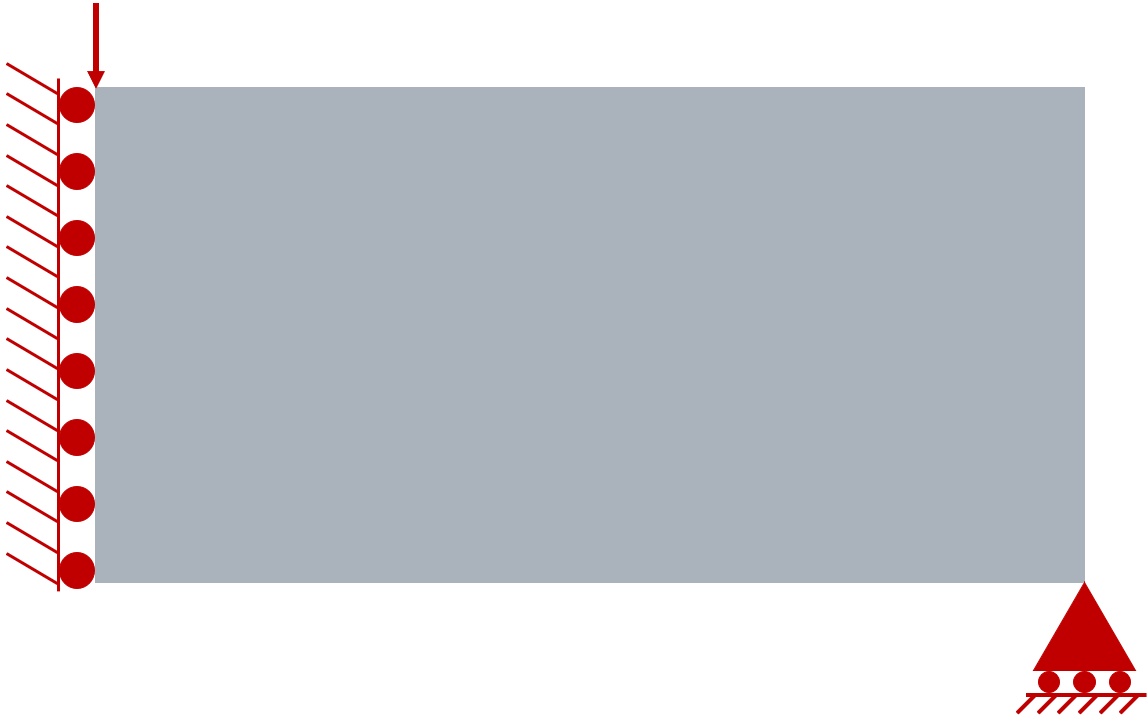}
		\subcaption{2D MBB beam: vertical load is applied on the top left corner, left edge is fixed horizontally and bottom right is fixed vertically.}
	\end{minipage}%
	\hspace{5mm}%
	\begin{minipage}[t]{.4\textwidth}
		\centering
		\includegraphics[width=\linewidth]{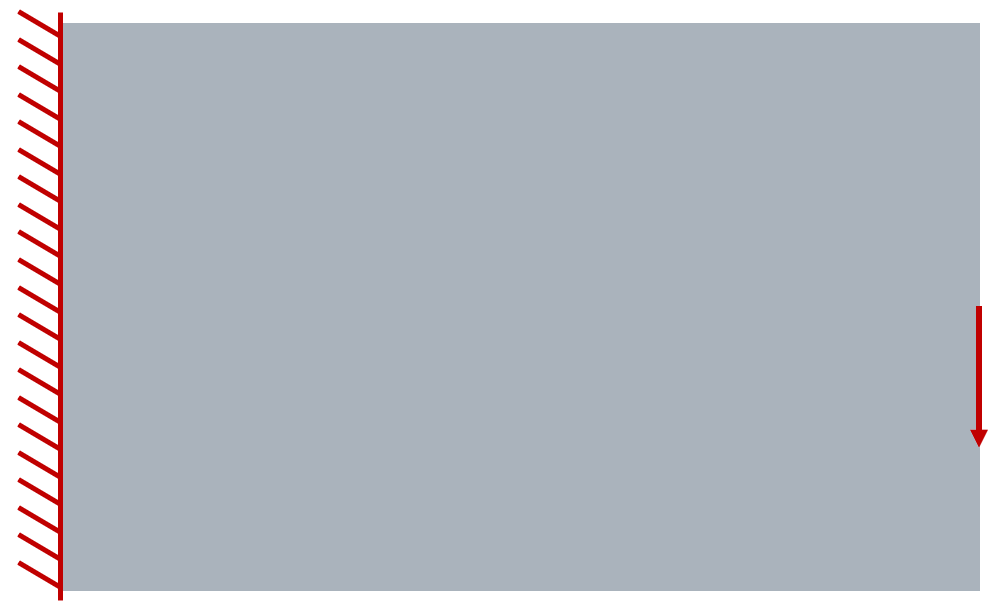}
		\subcaption{2D Cantilever beam: vertical load is applied on the middle of the right edge and the left edge is completely fixed.}
	\end{minipage}
	\caption{The boundary conditions of the 2D case studies used to illustrate the proposed method.}
	\label{fig:case_studies}
\end{figure}

\begin{figure}[h]
	\centering
	\begin{minipage}[t]{.4\textwidth}
		\centering
		\includegraphics[width=\linewidth]{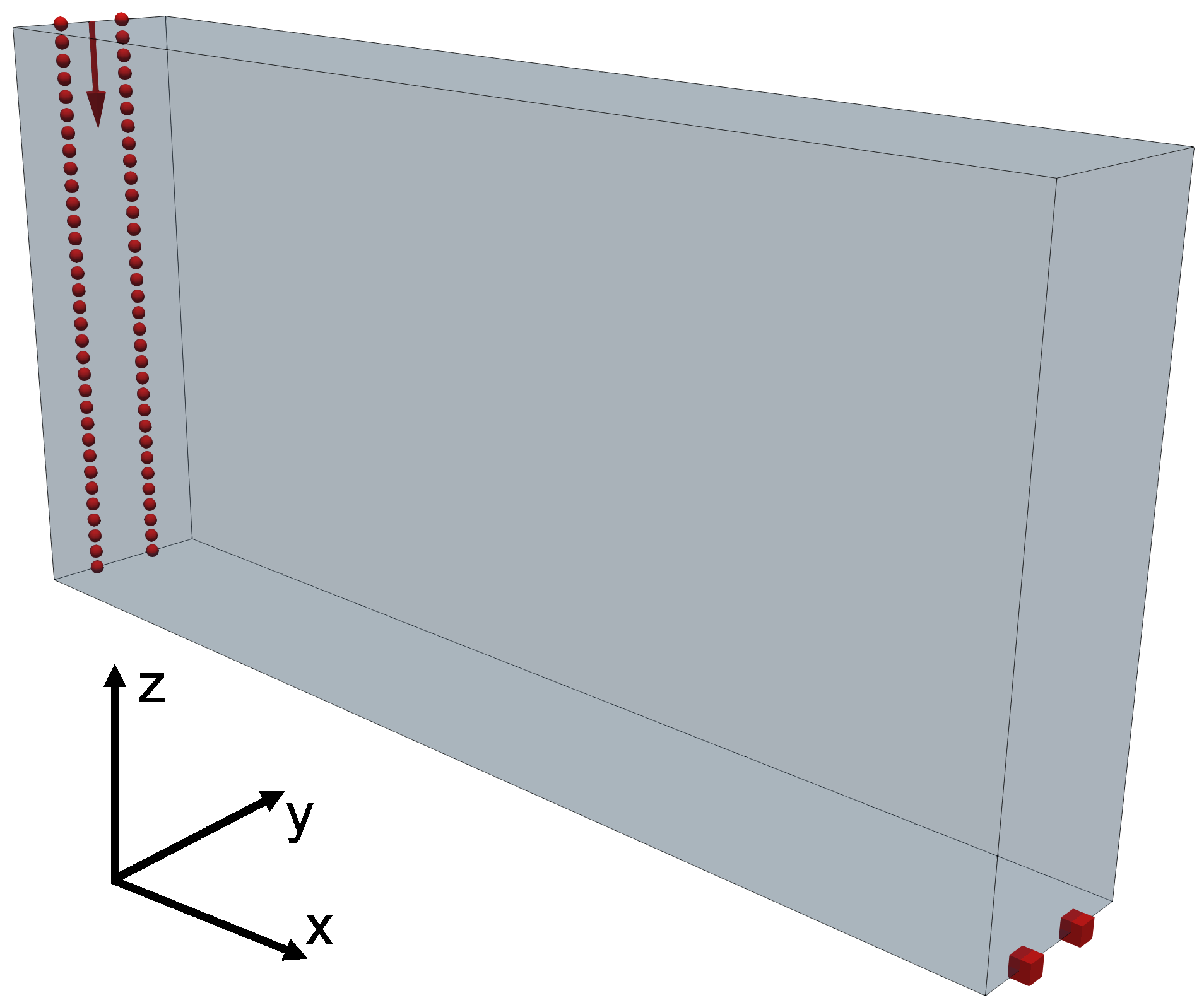}
		\subcaption{3D MBB beam: vertical load is applied on the top middle node denoted by an arrow, spheres denote nodes fixed in x-direction and cubes denote nodes fixed in y- and z-direction.}
	\end{minipage}%
	\hspace{5mm}%
	\begin{minipage}[t]{.38\textwidth}
		\centering
		\includegraphics[width=\linewidth]{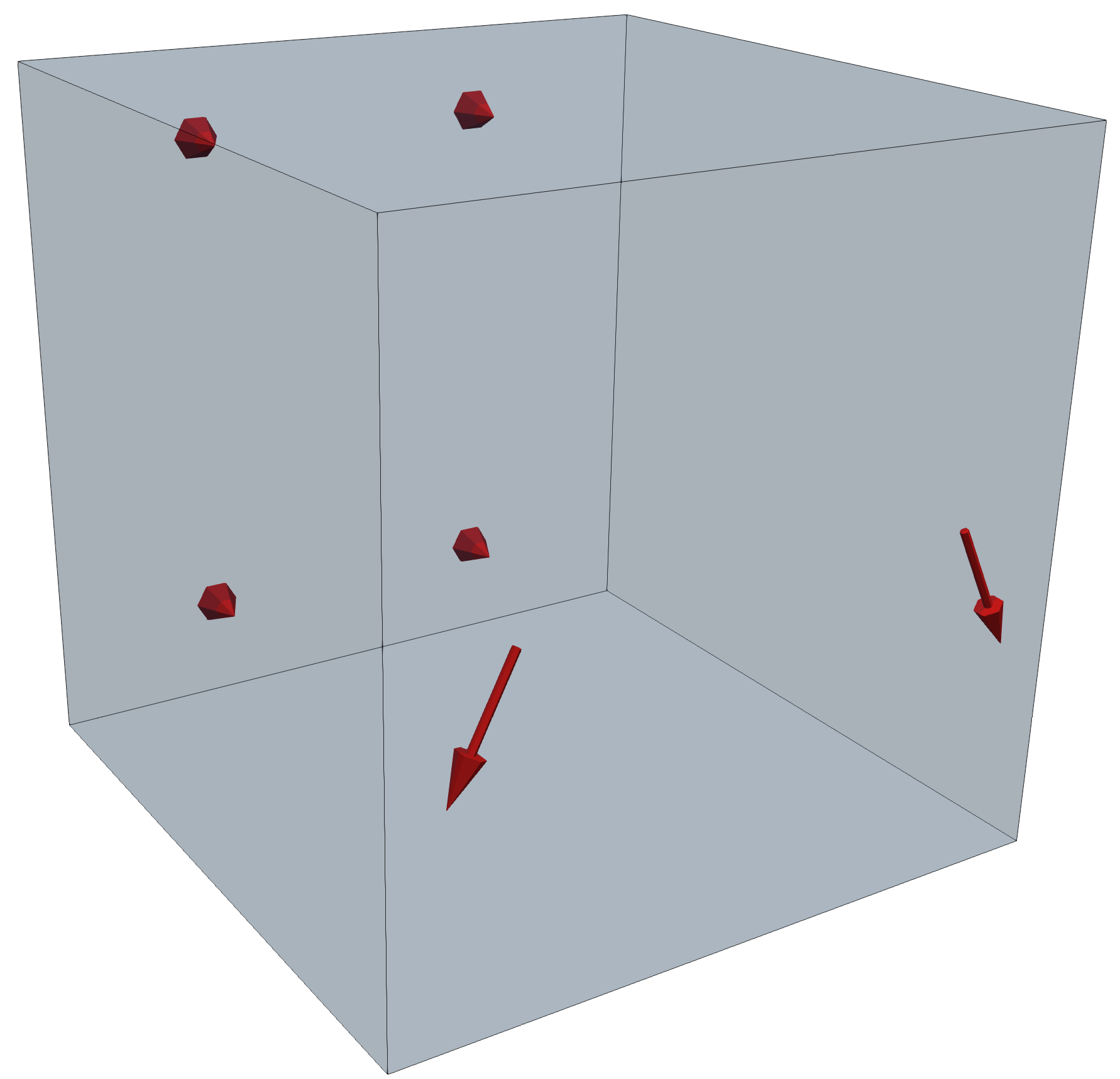}
		\subcaption{3D Cantilever beam: two loads are applied as denoted by the arrows and four nodes are completely fixed denoted by the cones.}
	\end{minipage}
	\caption{The boundary conditions of the 3D case studies used to illustrate the proposed method.}
	\label{fig:3Dcase_studies}
\end{figure}

Both 2D-cases and 3D-cases are presented in this section, with two examples each. One example is the MBB beam and the other is a cantilever beam, see respectively Figure \ref{fig:case_studies}a and \ref{fig:case_studies}b for the boundary conditions in 2D. The boundary conditions of the 3D case studies are shown in Figure \ref{fig:3Dcase_studies}a and \ref{fig:3Dcase_studies}b. The cases have been chosen such that complex topologies emerge, that provide a good test of the proposed method. The compliance of the examples is minimized given a certain maximum volume fraction. For simplicity, the same optimization problem is used in all stages to demonstrate the method. The Method of Moving Asymptotes (MMA) proposed by \citet{Svanberg1987method} is used as the optimization algorithm.

In all cases the TO of Stage 1 is performed using bi/trilinear shape functions. The shape optimization of Stage 3 is done using 2\textsuperscript{nd} order \textit{p}-FEM and one level quadtree or octree refinement near the boundary. The amount of Gauss points is 2$\times$2($\times$2) for Stage 1 and 3$\times$3($\times$3) in each integration cell for Stage 3 in 2D (or 3D). Other parameters used in the case studies are shown in Table \ref{tab:parameters}. As will be shown in Section~\ref{sec:performance}, this choice provides an attractive compromise between accuracy gain and computational cost.
\begin{table}[h]
	\footnotesize
	\centering
	\caption{Parameter values used for the case studies.}
	\label{tab:parameters}
	\begin{tabular}{|l|l|l|l|l|l|l|}
		\hline
		& $E$ & $\nu$ & $\rho_0$ & $p$ & $F_x,F_y(,F_z)$ & $s_{max}$ \\ \hline
		2D MBB& $1.0$ & $0.0$ & $1 \cdot 10^{-8}$ & $3$ & $0,-1$ & $0.5$  \\ \hline
		2D Canti& $1.0$ & $0.0$ & $1 \cdot 10^{-8}$ & $3$ & $0,-1$ & $0.5$  \\ \hline
		3D MBB& $1.0$ & $0.0$ & $1 \cdot 10^{-8}$ & $3$ & $0,0,-1$ & $0.3$  \\ \hline
		3D Canti& $1.0$ & $0.0$ & $1 \cdot 10^{-8}$ & $3$ & $0,\pm 0.5,-1$ & $0.3$  \\ \hline
	\end{tabular}
\end{table}

\begin{figure}[h]
	\centering
	\includegraphics[width=0.6\textwidth]{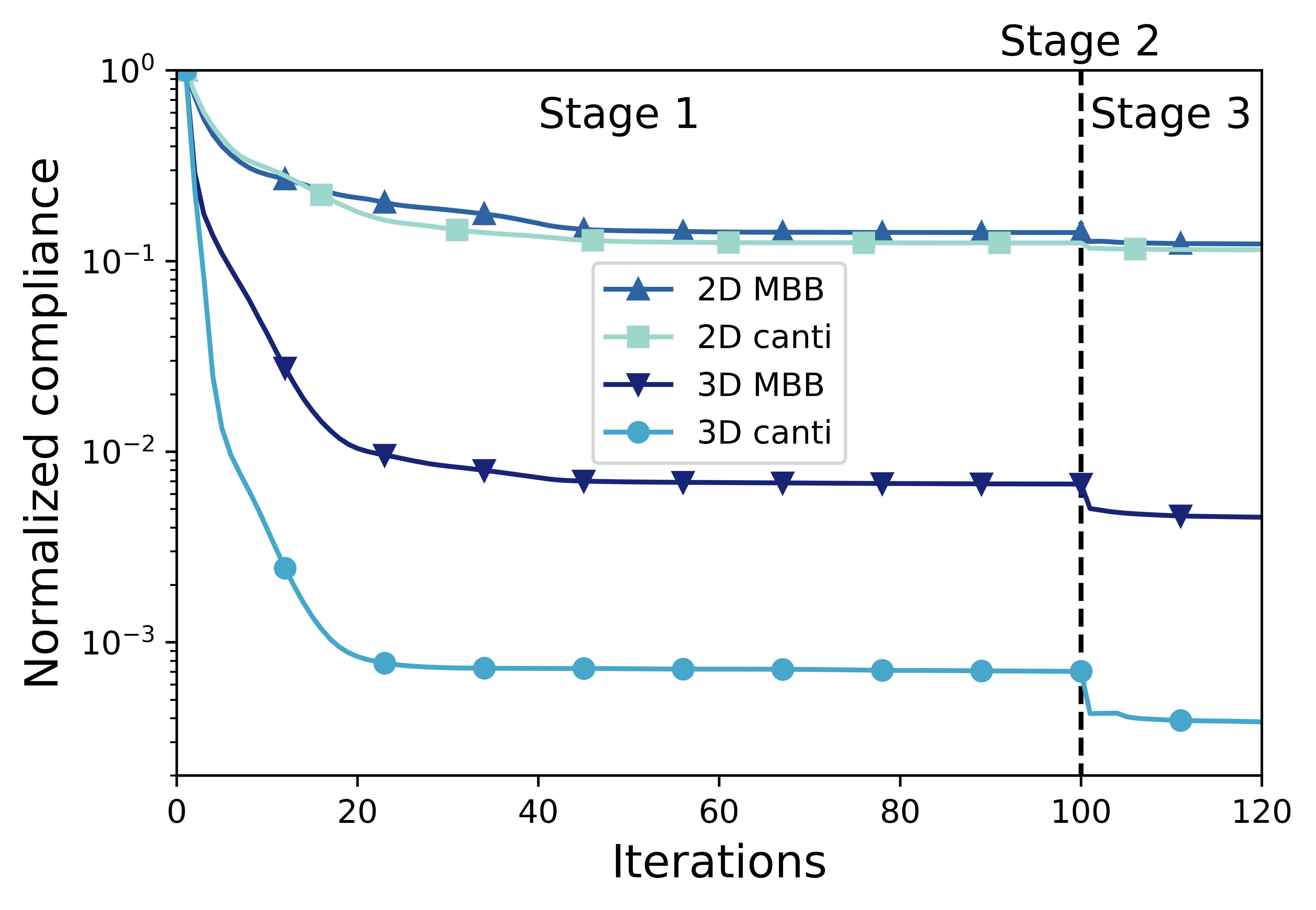}
	\captionsetup{width=0.6\textwidth}
	\caption{Convergence of the normalized compliance over the three design stages for the different case studies. The performance of the initial design of Stage 1 is used as a reference.}
	\label{fig:compl_conv}
\end{figure}

The results from the case studies are obtained using a fixed amount of 100 iterations for the topology optimization and two times 10 iterations for the shape optimization. The quadtree or octree integration band is reinitialized after the first 10 iterations of the shape optimization to make sure the boundary of the geometry does not move outside this region. A possible improvement of the algorithm could be to apply this reinitialization only when the evolving boundary reaches the edge of the refined integration band.
The convergence of the normalized compliance over the three design stages for the different case studies is shown in Figure \ref{fig:compl_conv}. The graph shows a steady improvement of the objective in each stage, and that the chosen number of iterations is adequate. The transformation from the final density-based design to the initial level-set description (Stage 2) already improves the compliance, mainly because inefficient intermediate density regions are removed while maintaining the maximum volume. Subsequently Stage 3 adds further improvement.

Next, we first present the numerical results of our post-processing method, followed by an analysis of speed and accuracy.

\subsection{2D-cases}
\begin{figure}[h]
	\centering
	\includegraphics[width=0.8\textwidth]{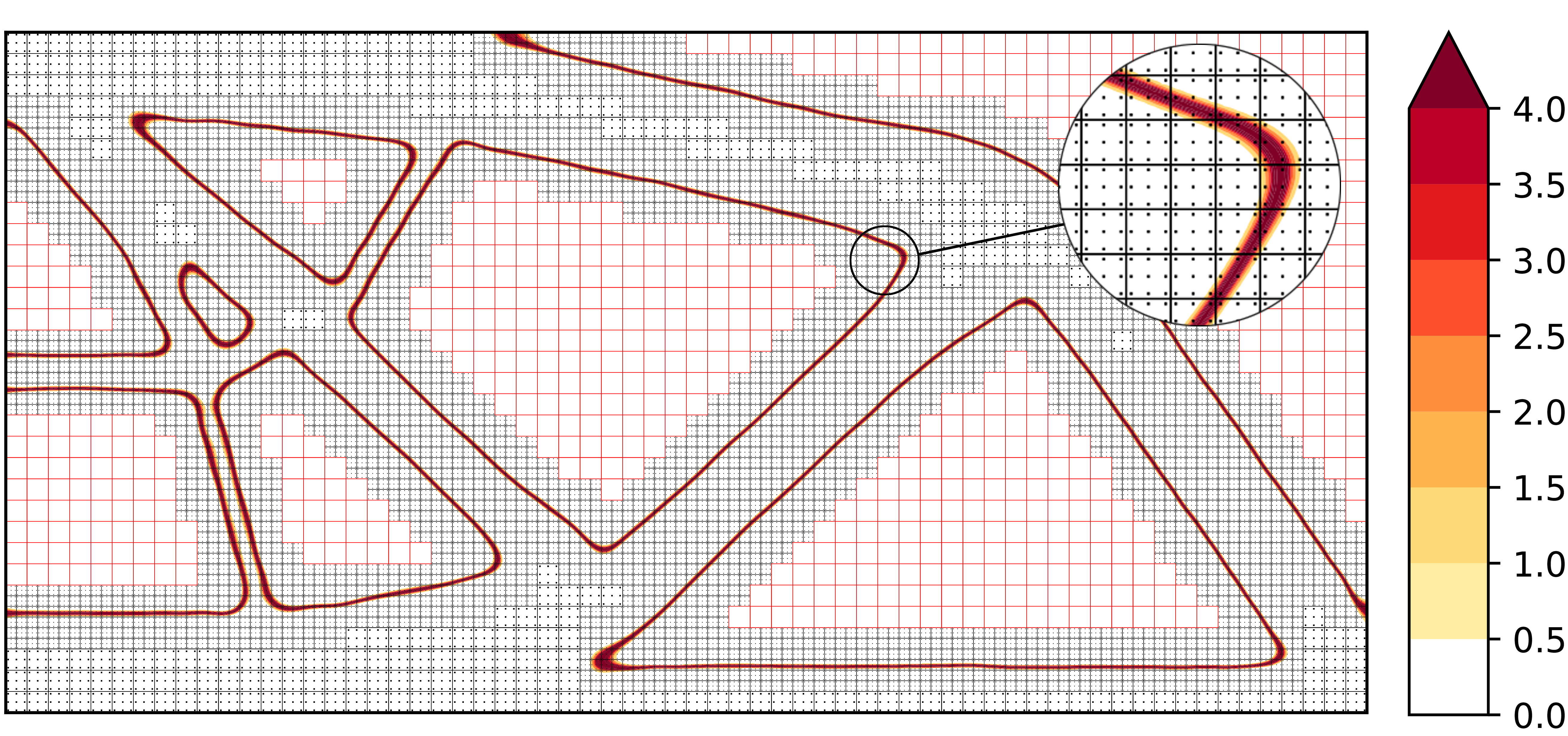}
	\captionsetup{width=0.8\textwidth}
	\caption{Sensitivity values $\partial{}C/\partial{}\mathbf{w}$ of the MBB beam after the shape optimization.}
	\label{fig:sens_MBB}
\end{figure}

\begin{figure}
	\centering
	\begin{minipage}[t]{.375\textheight}
		\centering
		\includegraphics[width=\linewidth]{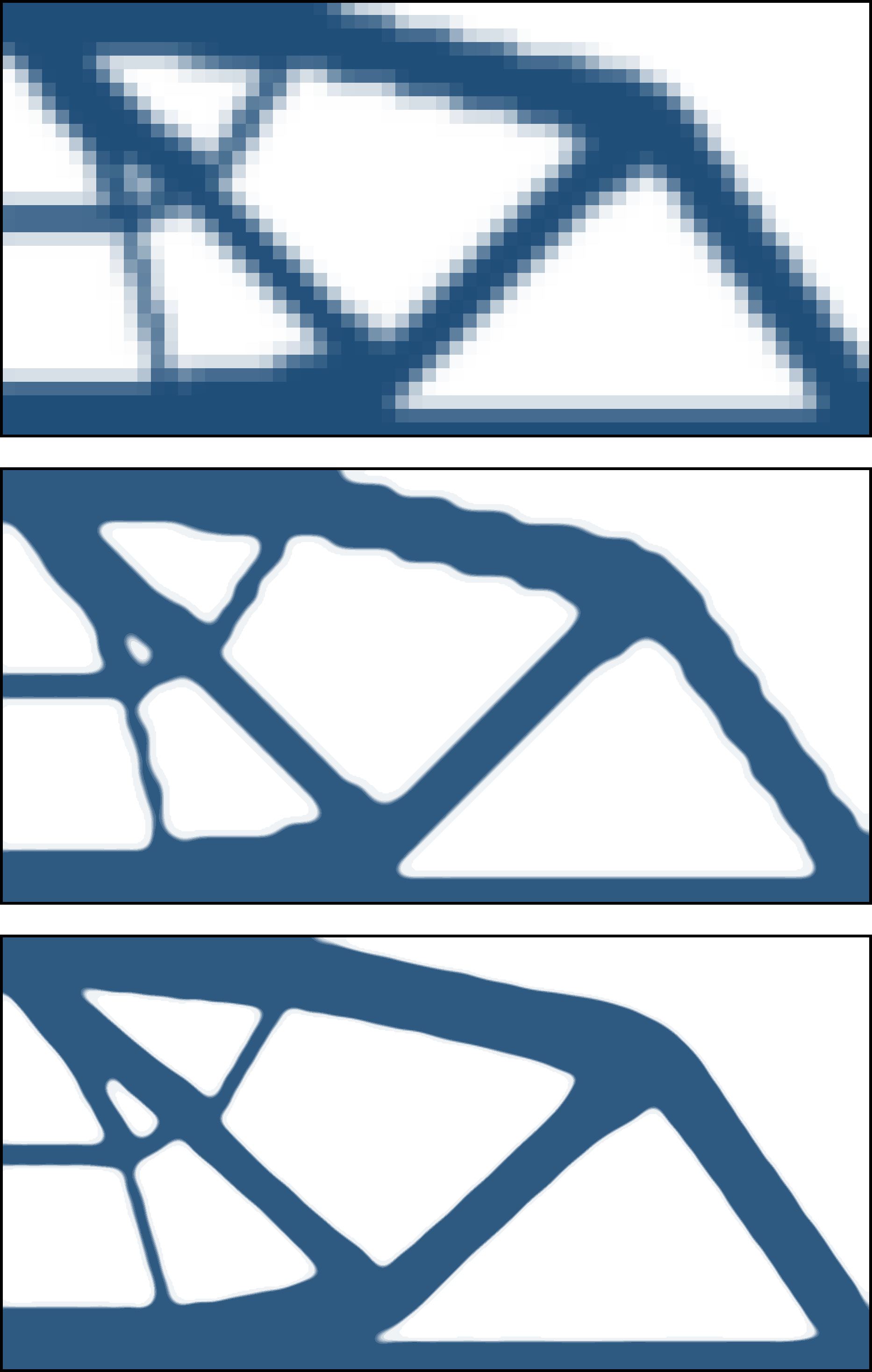}
		\subcaption{The MBB beam on a 64$\times$32 grid.}
	\end{minipage}%
	\hspace{6mm}%
	\begin{minipage}[t]{.2865\textheight}
		\centering
		\includegraphics[width=\linewidth]{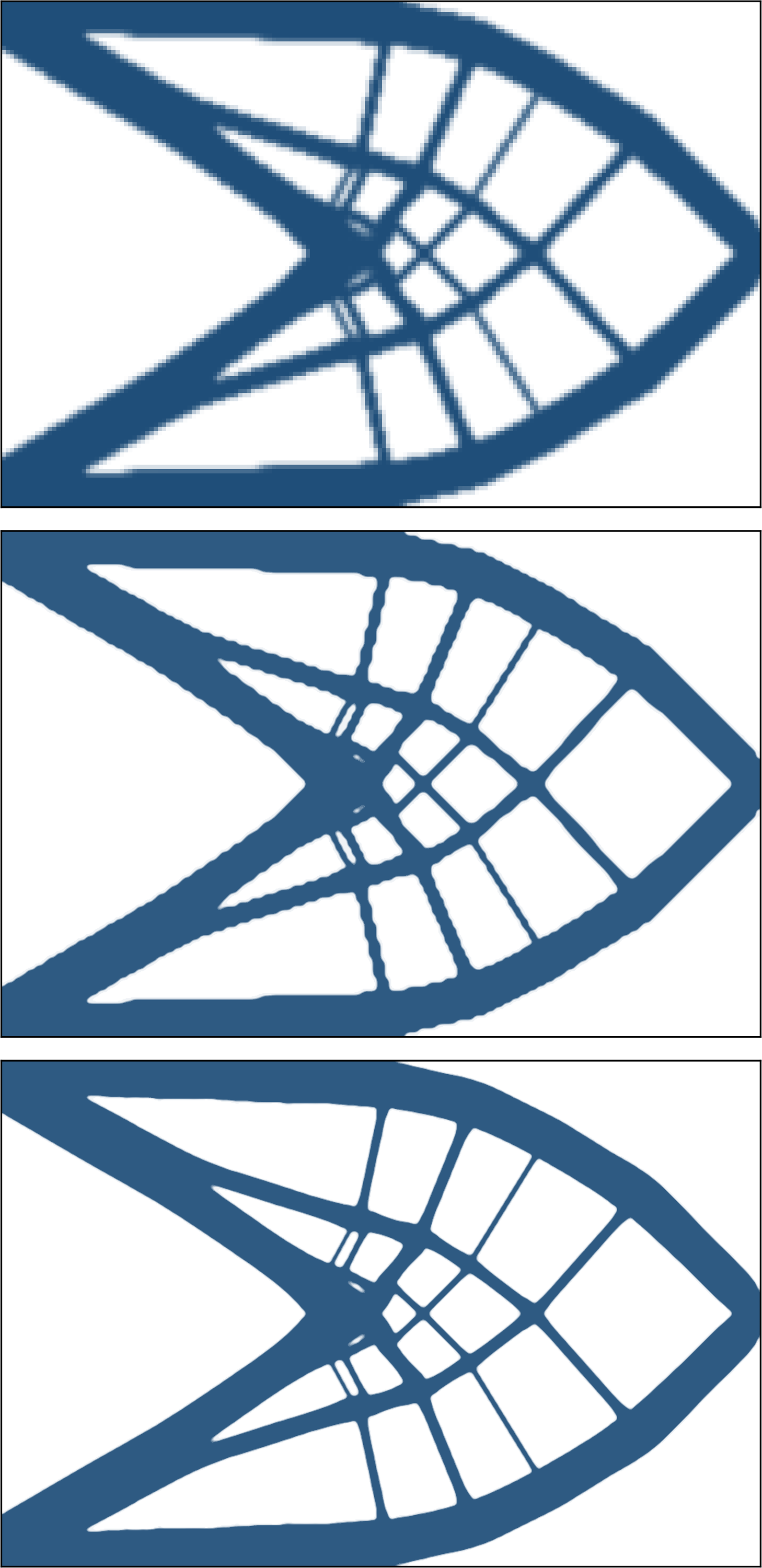}
		\subcaption{The cantilever beam on a 180$\times$120 grid.}
	\end{minipage}
	\caption{Results of the proposed three-stage structural design optimization process (topology optimization, geometry extraction and shape optimization) for the 2D case studies.}
	\label{fig:MBB_canti_overview}
\end{figure}

The MBB beam is optimized on a 64$\times$32 grid, allowing a volume fraction of 40\%. The results after each of the three different stages are shown in Figure \ref{fig:MBB_canti_overview}a. The geometry extracted from the TO results is already smooth and shows almost no intermediate densities. Stage 2 creates a direct smooth representation of the jagged boundaries, hence the wavy pattern for the extracted geometry. These patterns then disappear in the final design due to the shape optimization. In all cases, the volume constraint is satisfied after both Stage 2 and 3. It can be observed that a clear solid-void geometry with smooth boundaries is obtained, even when starting from a fairly coarse TO result with considerable artifacts.\\
It can be verified that the sensitivities are still defined close to the boundary for the final design after the shape optimization. Around the boundary there are always integration points having a non-zero sensitivity value, see Figure \ref{fig:sens_MBB}.

The cantilever beam is optimized on a finer 180$\times$120 grid, allowing a volume fraction of 35\%. This case illustrates the application of the procedure to a case with a more refined initial design. In spite of the higher resolution, jagged boundaries are still present in the TO result, as well as a blurred central region. The results after the three different stages are shown in Figure \ref{fig:MBB_canti_overview}b. This case study shows the capability of the proposed post-processing method to deal with intermediate densities. The blurred central spot is removed in the optimization stage and has disappeared in the final design obtained after Stage 3.

\subsection{3D-cases}
\begin{figure}[h!]
	\centering
	\begin{minipage}[t]{.35\textheight}
		\centering
		\includegraphics[width=\linewidth]{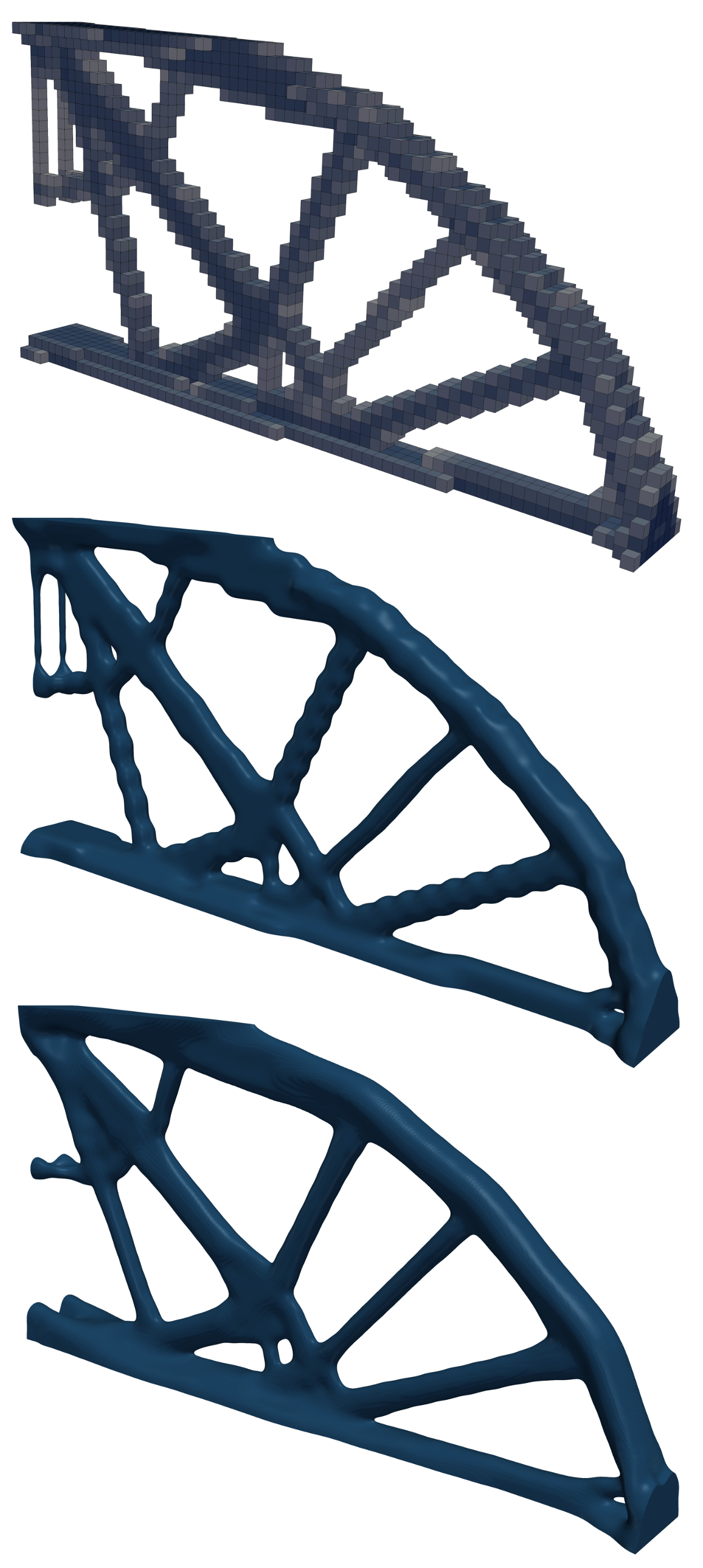}
		\subcaption{The 3D version of the MBB beam on a 64$\times$10$\times$32 grid.}
	\end{minipage}%
	\hspace{9mm}%
	\begin{minipage}[t]{.31\textheight}
		\centering
		\includegraphics[width=\linewidth]{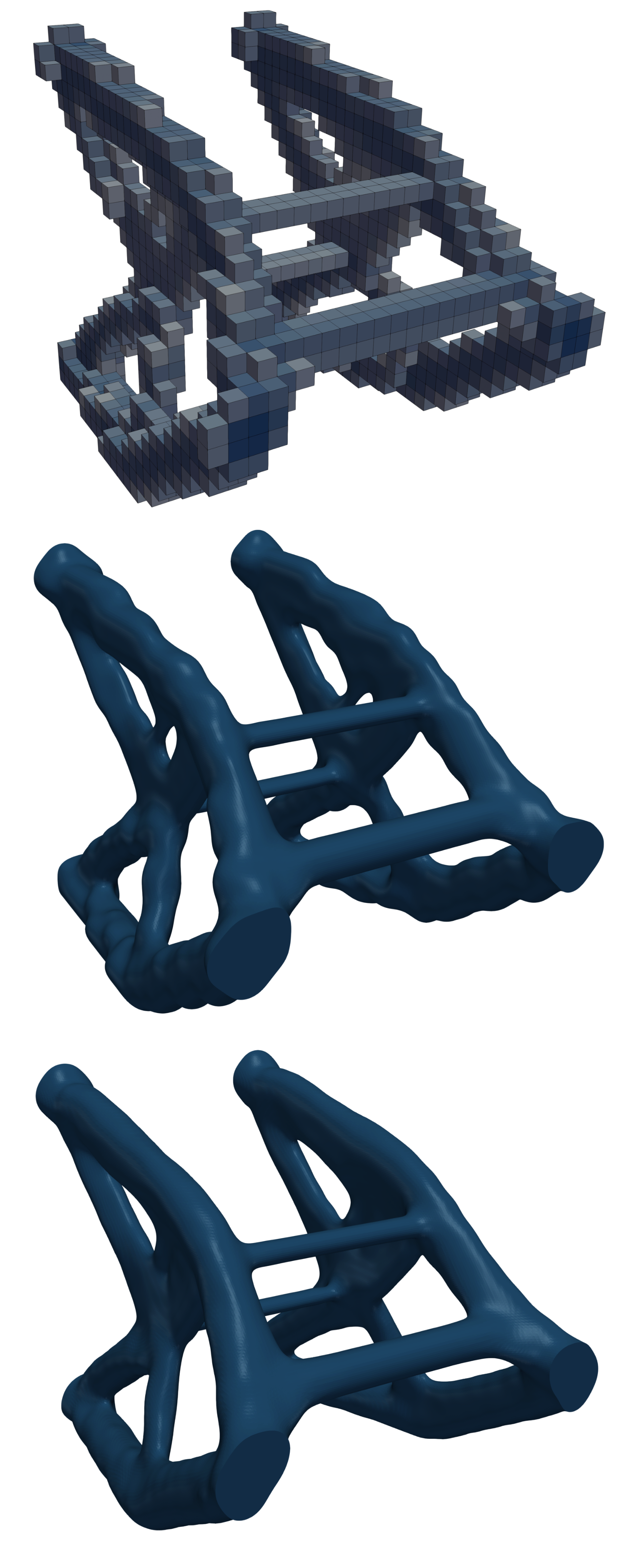}
		\subcaption{The 3D version of the cantilever beam, with alternative load placement and directions, on a 30$\times$30$\times$30 grid.}
	\end{minipage}
	\caption{Results of the proposed three-stage structural design optimization process (topology optimization, geometry extraction and shape optimization) for the 3D case studies.}
	\label{fig:3D_MBB_canti_overview}
\end{figure}

The MBB beam is also optimized in 3D on a 64$\times$10$\times$32 grid, allowing a volume fraction of 10\%. The 3D results are visualized by applying the Marching Cubes algorithm \citet{lorensen1987marching} to create a triangulated surface where the LSF is zero, which can be directly used to generate an STL input file for additive manufacturing, or processed further by other CAD tools. The MBB beams after the three different stages are shown in Figure \ref{fig:3D_MBB_canti_overview}a. This case study shows a change in the topology from Stage 2 to Stage 3. Two small members at the rear have disappeared due to the final optimization phase. This confirms that the level-set-based optimization in Stage 3 is not restricted to shape optimization, but can also handle topological changes.

The cantilever beam is also optimized in 3D but then on a 30$\times$30$\times$30 grid, allowing a volume fraction of 5\%. The loads are applied at two locations on the side plane in both vertical and transverse direction, in order to increase the complexity of the design to test the post-processing procedure. The results after the three different stages are shown in Figure \ref{fig:3D_MBB_canti_overview}b. Also here it can be seen that the proposed process generates a smooth, high-quality result from a relatively coarse 3D TO result, without any manual intervention.
\clearpage
\clearpage
\clearpage
\subsection{Performance}\label{sec:performance}
Next to the visual evaluation in the preceding sections, that confirms the smoothness and crispness of the generated designs, we also evaluate the performance of the proposed post-processing method on two additional aspects: speed and accuracy.

The computation time of the proposed post-processing method is difficult to quantify in a general sense, as it depends strongly on the chosen volume fraction, convergence criteria and desired accuracy. There also are possibilities for different trade-offs, in terms of resolution and iterations spent in Stage 1 (TO phase) versus effort spent on subsequent geometry extraction and shape optimization. No attempt has been made to optimize the total process in this regard.

For the considered example problems, computation times are listed in Table \ref{tab:comp_times}. We find the post-processing (Stage~2 and 3 combined) on average takes more time than the TO phase, excluding the 2D cantilever case. So, in the current implementation, the proposed post-processing method takes a higher computational effort than the (considerably coarser) TO stage. However, the employed Python code has not yet been optimized as the readability of the code was also considered important, and it leaves room for efficiency improvements. Moreover, comparing computation times of Stage~1 and Stages~2+3 is not a comparison on equal grounds. To achieve similar quality (smoothness, discreteness, analysis accuracy) using TO alone would require extensive refinement and considerably higher computational effort.
\begin{table}[h]
	\footnotesize
	\centering
	\caption{Computation times (s) for the case studies.}
	\label{tab:comp_times}
	\begin{tabular}{|l|r|r|r|r|}
		\hline
		Case & Stage 1 & Stage 2 & Stage 3 & Stage 2+3 \\ \hline
		2D MBB     & 20 & 1 & 22 & 53\% \\ \hline
		2D Canti   & 371 & 6 & 167 & 32\% \\ \hline
		3D MBB     & 1,203 & 53 & 3,108 & 72\% \\ \hline
		3D Canti   & 1,454 & 80 & 2,369 & 63\% \\ \hline
	\end{tabular}
\end{table}

\begin{figure}[h]
	\centering
	\includegraphics[width=0.5\textwidth]{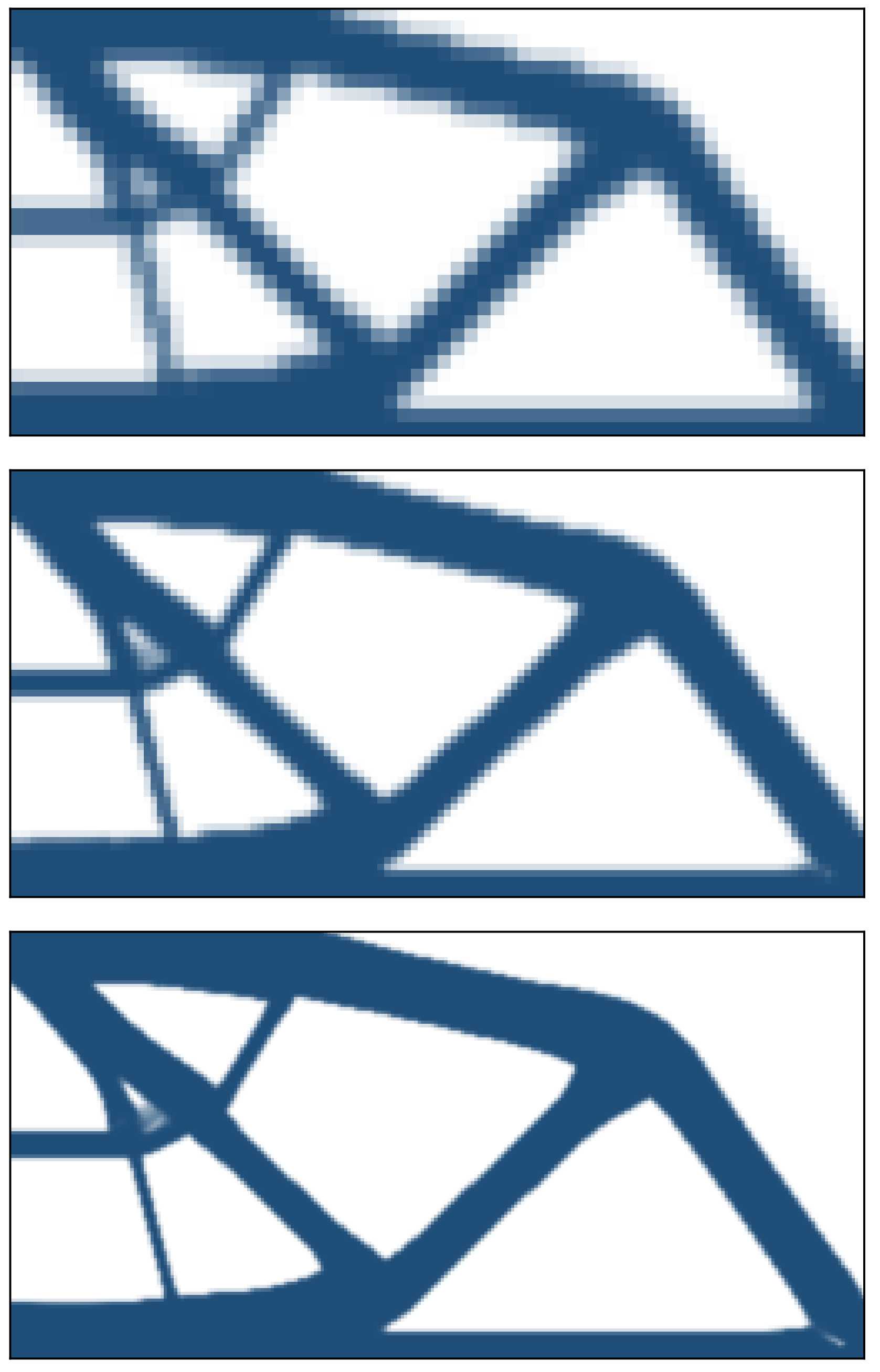}
	\captionsetup{width=0.6\textwidth}
	\caption{Improved resolution of the TO result of the MBB beam using an additional 20 iterations on a 64x32, an 128x64 and a 256x128 grid.}
	\label{fig:SIMP_refine}
\end{figure}

As an example to support this claim, the grid size is increased on the Stage~1 result of the MBB beam, and an additional twenty iterations of density-based topology optimization are performed to improve the resolution. The computation times are much higher as can be observed in Table \ref{tab:SIMP_refine}, especially compared to the proposed 3-staged method. Figure \ref{fig:SIMP_refine} shows the improved resolutions using the SIMP method. Even at a fourfold refinement level, at  \emph{over 10 times} the computational cost of the proposed post-processing method, the refined TO-only design still does not match the post-processed result shown in Figure \ref{fig:MBB_canti_overview}a.
\begin{table}[h]
	\footnotesize
	\centering
	\captionsetup{width=0.8\linewidth}
	\caption{Computation times for an additional twenty iterations of density-based topology optimization on different grid resolutions for the 2D MBB beam case study. For reference: Post-processing the base TO result (Stage 2+3) took 23 s.}
	\label{tab:SIMP_refine}
	\begin{tabular}{|l|r|}
		\hline
		Grid size  & Computation time (s) \\ \hline
		64x32   & 3.5 \\ \hline
		128x64  & 24.6 \\ \hline
		256x128 & 262.3 \\ \hline
	\end{tabular}
\end{table}

\begin{figure}[h!]
	\centering
	\begin{minipage}[t]{.475\textwidth}
		\centering
		\includegraphics[width=\linewidth]{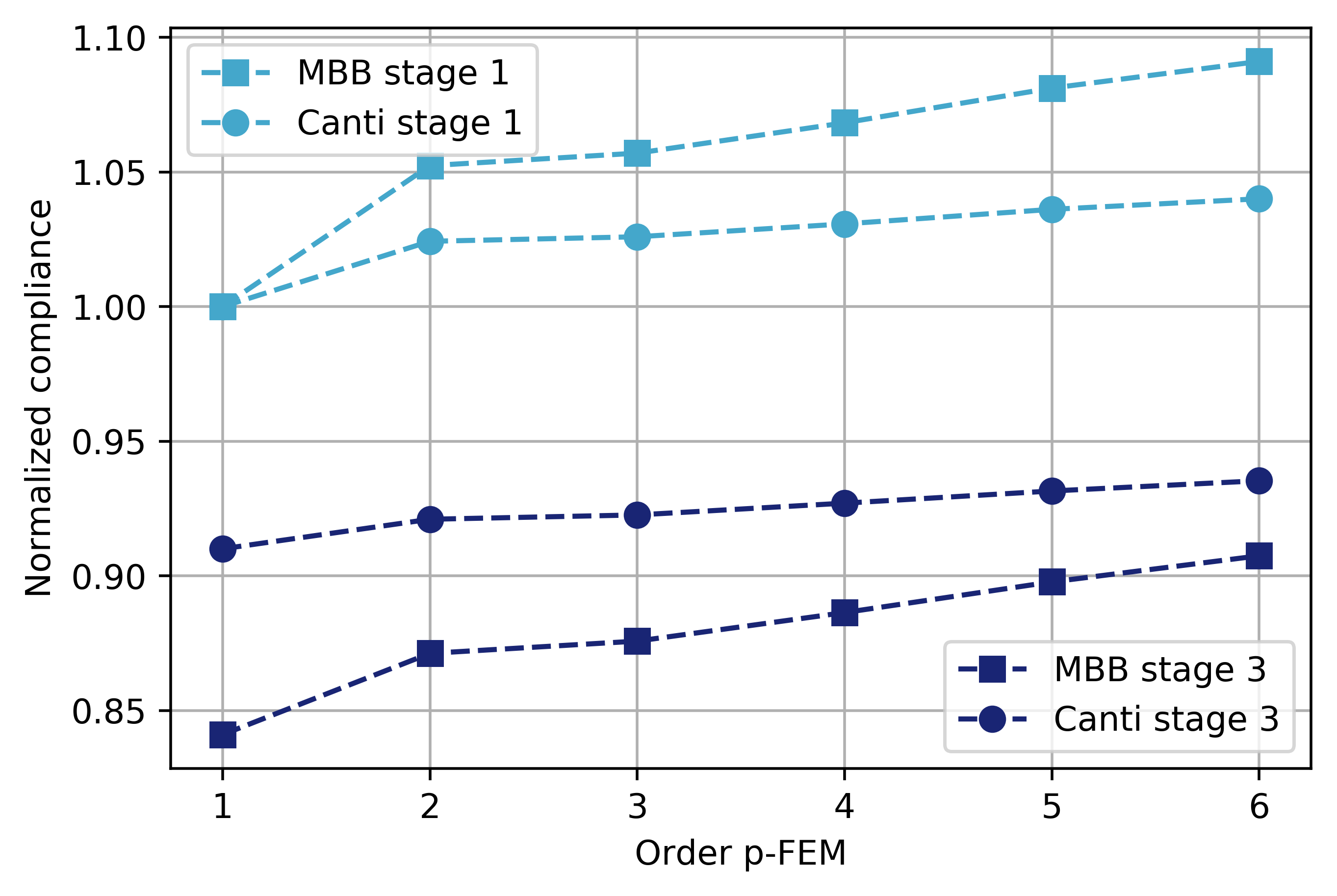}
		\subcaption{Normalized compliance for different order \textit{p}-FEM after Stage 1 and Stage 3 for both 2D case studies.}
	\end{minipage}%
	\hspace{4mm}%
	\begin{minipage}[t]{.475\textwidth}
		\centering
		\includegraphics[width=\linewidth]{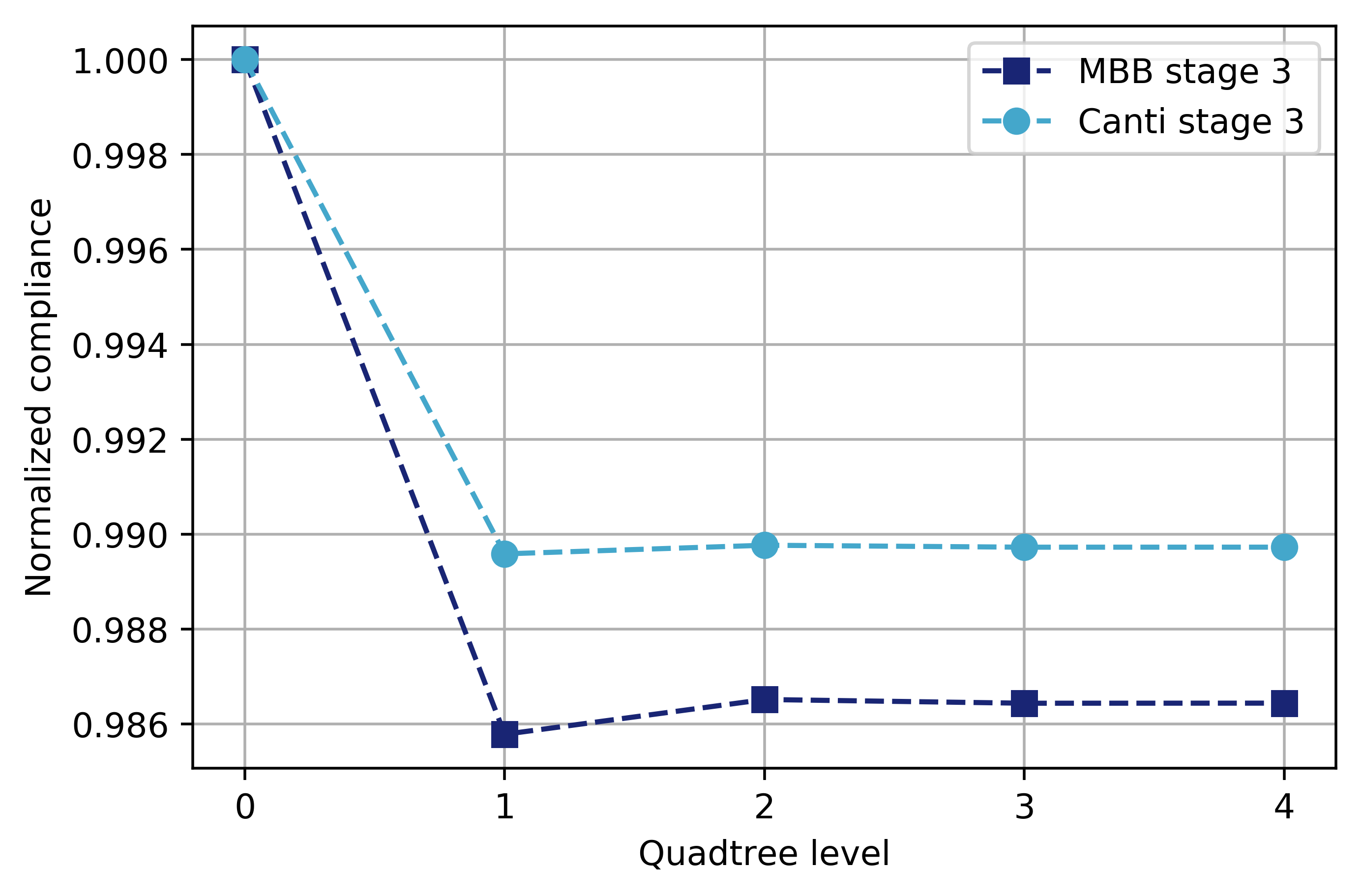}
		\subcaption{Normalized compliance for different levels quadtree refinement on the design after Stage 3 of both 2D case studies.}
	\end{minipage}
	\caption{Normalized compliance results for different FCM accuracy.}
	\label{fig:Estr_P_qt}
\end{figure}

The accuracy of the final result mainly depends on the accuracy of the structural analysis (i.e. order of \textit{p}-FEM and the number of quadtree levels). Figure \ref{fig:Estr_P_qt}a shows the normalized compliance for increasing \textit{p}-FEM order of the final designs obtained at Stage 1 and Stage 3 of the 2D case studies. The geometries based on an LSF are evaluated using four quadtree levels. All integration cells, also for the TO result, contain 8$\times$8 integration points to make sure there is no dominant integration error. For the Stage 3 results, $\kappa=25$ is used for the Heaviside function. The relatively highest increase in accuracy of the structural analysis is obtained by increasing from 1\textsuperscript{st} order \textit{p}-FEM to 2\textsuperscript{nd} order.

Note that $p$-FEM orders of 6 or higher would be needed to reach numerical convergence. The associated computational cost is not considered practical for the design stage. For this reason, we find that $p$=2 provides a good compromise between cost and accuracy. Note also that all Stage 3 cases outperform the Stage 1 results, i.e. the post-processing process improves the performance of the designs by 10-15\%, as can be observed in Figure \ref{fig:compl_conv} as well.\\
Figure \ref{fig:Estr_P_qt}b shows the normalized compliance for increasing levels of quadtree refinement of the final designs obtained at Stage 3 of the 2D case studies. The use of more than one quadtree level, in combination with 2\textsuperscript{nd} order \textit{p}-FEM and 3$\times$3 Gauss integration points, does not influence the compliance of the analysed designs significantly, justifying the use of only one quadtree level for the case studies.
\clearpage
\section{Conclusion}
In this paper, we have presented a method for automated post-processing of both 2D and 3D density-based TO results using a fully integrated level-set-based shape optimization, in order to efficiently improve the quality of TO results and to form a convenient bridge to other CAD tools. The aim was to obtain a structural design optimization process which results in optimized, smooth and crisp geometries without any manual labour. Based on the numerical examples, it can be concluded that this goal has been achieved. In contrast to existing approaches, the method extends naturally to 3D and produces optimized designs, instead of only performing a smoothing step for visual appearance. The chosen level-set-based formulation actually allows for topological changes in the final optimization, which enables the removal of inefficient structural features.\\
Our method exploits \textit{p}-refinement to avoid remeshing, and second order shape functions combined with a single quadtree/octree refinement proved sufficient to obtain accurate shape optimization results. This is linked to the way the maximum slope of the employed LSF is controlled, by bounding the weights of the involved RBFs. A rigorous procedure to set suitable weight bounds forms another contribution of this paper.

%Recommendations
As future work, it will be of interest to test this approach on other TO problems involving different objectives and constraints. The presented approach and the associated sensitivity analysis can easily be extended to other types of optimization problems, e.g. the use of stress constraints as done in \citet{cai2014stress} and \citet{cai2015stress}. Also, it was found that choices can be made whether to spend most computational effort in the initial TO stage, or in the final shape optimization. How to optimally balance these two steps is also still to be investigated. In the considered examples, the computational time required for the proposed post-processing steps is more on average than that used in the TO stage, but efficiency improvement potential remains for the current implementation. Compared to improving designs through manual post-processing or higher resolution TO, the proposed fully automated procedure is certainly an attractive option. The developed Python code for the 2D case studies is made available and can be found on Github via this \href{https://bit.ly/2MVDciG}{link}.

\section*{Acknowledgements}
We gratefully acknowledge the contribution of J.C. Bakker, M. Barbera, N.D. Bos and S.J. van Elsloo to the Python script written for this research.\\
We would like to thank Femto Engineering as well for their partnership in this research.

%\bibliographystyle{abbrvnat}  %unsrt , apalike , plain, abbrv
%\bibliography{./library.bbl}  

\end{document}